\documentclass[letterpaper,11pt]{article}
\pdfoutput=1 

\usepackage{jcappub} 

\usepackage[T1]{fontenc} 

\usepackage{booktabs}

\def\beq{\begin{equation}}
\def\eeq{\end{equation}}
\def\bp{\boldsymbol{p}}

\def\bk{\boldsymbol{k}}

\def\bx{\boldsymbol{x}}

\newcommand*\diff{\mathop{}\!\mathrm{d}}

\newcommand{\CL}{\texttt{$\mathcal{C}\text{osmo}\mathcal{L}\text{attice}$}}

\title{Impact of Dark Sector Preheating on CMB Observables}

\author{Marcos A. G. Garcia}
\author{and Aline Pereyra-Flores}

\affiliation{Departamento de F\'isica Te\'orica, Instituto de F\'isica,\\
Universidad Nacional Aut\'onoma de M\'exico, Ciudad de M\'exico C.P. 04510, Mexico}

\emailAdd{marcos.garcia@fisica.unam.mx}
\emailAdd{pereyrafa@estudiantes.fisica.unam.mx}

\abstract{The prediction of a nearly scale-invariant spectrum of curvature and tensor fluctuations is among the main features of cosmic inflation. The current measurements of the primordial fluctuations in the cosmic microwave background (CMB) provide tight constraints on the amplitude of the scalar and tensor spectra, and the scalar tilt. However, the precise connection between these observables and a given inflationary model, depends on the expansion history between the end of inflation and the beginning of the radiation dominated era, which corresponds to the reheating epoch. This mapping between horizon exit and reentry of fluctuations, parametrized by the number of $e$-folds $N_*$,  can therefore be affected by the presence of a transient epoch of non-perturbative particle production during reheating ({\em preheating}). Using a combination of perturbative and lattice computations, we quantify the impact of preheating in a non-equilibrated dark matter sector on the CMB observables, under the assumption of a simultaneous perturbative decay of the inflaton into Standard Model particles. Combined with structure formation constraints, this allows us to impose stringent bounds on the post-inflationary reheating temperature.
 }

\begin{document}
\maketitle
\flushbottom

\section{Introduction}\label{Intro}

The successful solution of the classical problems of the Standard Big Bang cosmological scenario (the flatness, horizon, and monopole problems), and the prediction of a near scale-invariant spectrum of primordial fluctuations, are among the main features of the inflationary paradigm~\cite{Olive:1989nu,Linde:1990flp,Lyth:1998xn,Linde:2000kn}. In its arguably simplest realization, inflation is realized by the slow-roll of a real, scalar field known as the inflaton, down a flat potential. Quantum fluctuations in the inflaton field and the spacetime metric are also stretched by the expansion beyond the horizon scale, turned into curvature and tensor fluctuations. Their primordial spectra leave an imprint in the form of temperature and polarization anisotropies of the cosmic microwave background (CMB). Their measurement and characterization by WMAP, {\em Planck}, and BICEP/{\em Keck} provide now a stringent bound at the pivot scale $k_*=0.05\,{\rm Mpc}^{-1}$, on the ratio of the tensor power spectrum relative to the scalar one, $r< 0.036$, at the 95\% CL. Moreover, the tilt of the scalar spectrum lies in the range $0.958<n_s<0.975$ for $r=0.004$ at the 95\% CL~\cite{BICEP:2021xfz}. These values for the CMB observables can be comfortably accommodated in scenarios with asymptotically flat inflaton potentials, among which the $R+R^2$ Starobinsky model can be singled out in particular~\cite{Starobinsky:1980te}. 

An additional feature of slow-roll inflation is the automatic graceful exit from accelerated expansion, as the field reaches the minimum of its potential, and commences to coherently oscillate about it. The dissipation of the energy stored in these oscillations into a hot, thermal plasma of relativistic particles reheats the universe. Generically, this reheating not only populates the Standard Model (SM) degrees of freedom, but also the degrees of freedom that make up the dark sector of the universe. At the perturbative level, the dissipation process is modeled as the quantum mechanical decay of inflaton quanta into elementary particles, with a decay rate determined by the adiabatic transition amplitude, ignoring the short time-scale oscillation of the inflaton field~\cite{Abbott:1982hn,Shtanov:1994ce,Kofman:1994rk,Ichikawa:2008ne,Amin:2014eta,Kainulainen:2016vzv,Garcia:2020wiy,Garcia:2020eof}. 

There exist, however, scenarios in which the high frequency oscillation of the inflaton is the primary driver for particle production. In particular, integer-spin light fields coupled to the inflaton can undergo an exponential growth of their fluctuations, due to the resonant amplification of their field modes originated from a nonadiabatic change in their effective masses. This nonperturbative particle production mechanism is known as {\em preheating}, and can significantly alter the post-inflationary expansion history of the universe~\cite{Shtanov:1994ce,Dolgov:1989us,Traschen:1990sw,Boyanovsky:1995ud,Yoshimura:1995gc,Kofman:1997yn}. When the amplification of fluctuations is strong, it can backreact into the classical, coherent inflaton condensate via mode-mode couplings, sourcing large gradients and eventually leading to its fragmentation~\cite{Amin:2014eta,Garcia-Bellido:2002fsq,Felder:2006cc,Frolov:2010sz,Garcia:2021iag,Figueroa:2016wxr}. Although this non-perturbative production of particles does not lead in general to a complete reheating and a radiation dominated era, it can play a role in the efficiency of the decay of the inflaton into light degrees of freedom in the early and late  stages of reheating, modifying the duration of the reheating process~\cite{Garcia:2023eol}. Importantly, despite the fact that super-horizon fluctuations are generically impervious to these complex dynamics~\cite{Jedamzik:1999um,Ivanov:1999hz,Liddle:1999hq}, the precise connection between a given observable scale and its horizon-exit during inflation depends on the duration of reheating and the corresponding change in the total energy density of the universe~\cite{Liddle:2003as,Martin:2006rs,Martin:2010kz,Mielczarek:2010ag,Dai:2014jja,Martin:2014nya,Cook:2015vqa,Ellis:2015pla,Drewes:2015coa,Ellis:2021kad,Drewes:2022nhu,Drewes:2023bbs}. Therefore, the presence of preheating, on top of a dissipative, perturbative decay of the inflaton, can add to the uncertainty that reheating embodies on the connection between a given inflationary model and its corresponding CMB predictions~\cite{Podolsky:2005bw,Lozanov:2016hid,Lozanov:2017hjm,Jiang:2018uce,Antusch:2020iyq}.

As studied by one of the authors in previous works, preheating provides an efficient out-of-equilibrium mechanism for generating the relic dark matter abundance of the universe~\cite{Garcia:2021iag,Garcia:2022vwm,Garcia:2023awt}. In this paper we explore the impact of dark sector preheating on the CMB observables $n_s$ and $r$, in the case when the inflaton oscillates about a quadratic minimum of its potential. For definiteness we consider a scalar dark matter (DM) candidate, minimally coupled to gravity, which interacts directly only with a Starobinsky-like inflaton sector. We numerically track the evolution during reheating of the inflaton field, the dark matter fluctuations, and the relativistic thermal plasma which contains the Standard Model particles, which we assume to be produced perturbatively. For large inflaton-DM couplings, we track the instantaneous energy density of the plasma before, during, and after the backreaction regime, during which the oscillating inflaton condensate is fragmented in favor of an inhomogeneous collection of free particles. As a result, we obtain semi-analytical estimates of the change in the number of $e$-folds, the scalar tilt, the tensor-to-scalar ratio, and the DM relic abundance, in terms of the depletion of the comoving inflaton energy density due to backreaction. From these parameters, we derive strong constraints on the allowed couplings between the inflaton and the visible sector fields.

This paper is organized as follows. In Section~\ref{sec:inflation} we revisit the analytical and numerical approximations necessary to estimate to high accuracy the CMB observables for the Starobinsky model of inflation, with emphasis on their dependence on the reheating epoch. Section~\ref{sec:preheating} is devoted to the discussion of the parametric excitation of dark degrees of freedom due to their coupling to the coherently oscillating inflaton during reheating. We present a numerical exploration of the nonlinear backreaction regime, and the resulting fragmentation of the inflaton condensate. Our main results are discussed in Section~\ref{sec:signatures}. In Section~\ref{sec:reheatingfrag} we find a parametrization of the effect of strong DM preheating on the evolution of the inflaton and radiation energy densities during reheating. In Section~\ref{sec:DM} we compute the DM relic abundance for strong preheating, and determine the effect of the backreaction onto the inflaton field. Section~\ref{sec:CMB} is devoted to the precise computation of the CMB observables, and to mapping the current observational constraints onto the effective strength of the inflaton-matter couplings. We present a summary and our conclusions in Section~\ref{sec:conclusions}.

\section{Inflationary dynamics}\label{sec:inflation}

The field content of our study consists of an inflaton field $\phi$, a sterile scalar dark matter field $\chi$, and the Standard Model degrees of freedom, relativistic at early times, to which we will collectively refer in terms of their total energy density $\rho_R$. We write the corresponding action as
\beq
\label{eq:action}
\mathcal{S} \;=\; \int \diff ^4x\,\sqrt{-g} \left[\frac{1}{2}(\partial_{\mu}\phi)^2 - V(\phi) + \frac{1}{2}(\partial_{\mu}\chi)^2 - \frac{1}{2}m_{\chi}^2\chi^2 - \frac{1}{2}\sigma \phi^2\chi^2 + \mathcal{L}_{\rm SM}\right] \, .
\eeq
Here $g \equiv \det(g_{\mu \nu})$ denotes the metric determinant of a flat Friedmann-Robertson-Walker metric with scale factor $a$, $m_{\chi}$ is the DM mass, $\sigma$ is the dimensionless inflaton-DM coupling, $\mathcal{L}_{\rm SM}$ contains all the SM terms, including their coupling to the inflaton, and $V(\phi)$ denotes the inflaton potential. For definiteness we choose this potential as the Starobinsky potential,
\begin{align}\label{eq:staropot}
V(\phi) \;&=\; \frac{3}{4}m_{\phi}^2M_P^2\left( 1 - e^{-\sqrt{\frac{2}{3}}\frac{\phi}{M_P}}\right)^2\,,
\end{align}
where $M_P = 1/\sqrt{8\pi G_N}\simeq 2.435\times 10^{18}\,{\rm GeV}$ denotes the reduced Planck mass. In its original inception, this model arises as the effective scalar-tensor theory in the Einstein frame of the $R^2$ generalization of the Einstein-Hilbert action~\cite{Starobinsky:1980te}. The model also naturally appears in no-scale supergravity inflation models, in which a singularity in the kinetic term of the inflaton is manifested as the asymptotically flat potential (\ref{eq:staropot}) once the field is translated into its canonically normalized counterpart~\cite{Ellis:2013xoa,Ellis:2013nxa,Ellis:2014gxa,Ellis:2015xna,Ellis:2020lnc}. 

Disregarding the coupling of the inflaton to other fields during inflation, variation of the action (\ref{eq:action}) with respect to the homogeneous inflaton field and metric yields the equations of motion
\begin{align}\label{eq:KGeq}
\ddot{\phi} + 3H\dot{\phi} + \partial_{\phi}V \;&=\; 0\,,\\ \label{eq:Feq}
\frac{1}{2}\dot{\phi}^2 + V \;&=\; 3H^2 M_P^2 \,,
\end{align}
where dots denote derivatives with respect to cosmic time, and $H=\dot{a}/a$ is the Hubble parameter. The solution of this system describes the slow-roll of the inflaton field as it rolls down the potential (\ref{eq:staropot}), and the quasi-exponential growth of the scale factor $a$. Although the numerical solution of Eqs.~(\ref{eq:KGeq})-(\ref{eq:Feq}) is a straightforward exercise, analytical approximations for the amount of expansion and the inflationary observables can be obtained via the slow-roll approximation, which we review for the Starobinsky potential below. 

\subsection{Slow-roll approximation}

Near the minimum, for $\phi\ll M_P$, the potential is quadratic, $V(\phi)\simeq \frac{1}{2}m_{\phi}^2\phi^2$, and thus the parameter $m_{\phi}$ corresponds to the inflaton mass. Its value is determined by the amplitude of the curvature power spectrum, $A_{S}\simeq 2.1\times 10^{-9}$, evaluated at the horizon exit time of the {\em Planck} pivot scale $k_P=0.05\,{\rm Mpc}^{-1}$~\cite{Planck:2018vyg,Planck:2018jri}. The connection can be determined by means of the slow-roll approximation. This approximation is valid when the slow-roll parameters
\beq
\epsilon \;\equiv\; \frac{1}{2}M_P^2\left(\frac{V_{\phi}}{V}\right)^2\,,\qquad \eta\;\equiv\; M_P^2\left(\frac{V_{\phi\phi}}{V}\right)\,,
\eeq
are both $\ll 1$. In terms of these parameters, the amplitude of the scalar spectrum can be approximated as
\beq\label{eq:As}
A_{S*} \;\simeq\; \frac{V_*}{24\pi^2\epsilon_* M_P^4}\,,
\eeq
where the star denotes evaluation at the horizon crossing time of the scale $k_*$. The field value at this time can be computed in terms of the number of $e$-folds to the end of inflation 
\beq
N_* \;\simeq\; \int_{\phi_{\rm end}}^{\phi_*} \frac{d\phi}{\sqrt{2\epsilon} M_P}\,,
\eeq
yielding
\beq\label{eq:phistar}
\phi_{*} \;\simeq\; M_P\sqrt{\frac{3}{2}}\left[ 1+ \frac{3}{4N_*-3} \right] \ln\left( \frac{4N_*}{3} + e^{\sqrt{\frac{2}{3}}\frac{\phi_{\rm end}}{M_P}} - \sqrt{\frac{2}{3}}\frac{\phi_{\rm end}}{M_P}\right)\,,
\eeq
where 
\beq\label{eq:phiend}
\phi_{\rm end} \;\simeq\;  0.615\,M_P\,,
\eeq
is the inflaton field value at the end of inflation, when $\ddot{a}=0$, or equivalently $\dot{\phi}_{\rm end}^2=V(\phi_{\rm end})$~\cite{Ellis:2015pla,Ellis:2021kad}. Substitution into (\ref{eq:As}) gives the following mass normalization,
\beq\label{eq:mphi}
m_{\phi} \;\simeq\; \frac{2\pi M_P \sqrt{6A_{S*}}}{N_*} \;\simeq\; 3\times 10^{13}\,{\rm GeV}\,\left(\frac{55}{N_*}\right)\,.
\eeq

The scalar tilt $n_s$, and the tensor-to-scalar ratio $r$, can be determined to leading order from the slow-roll parameters as follows~\cite{Bezrukov:2007ep},
\begin{align}\label{eq:nsN}
n_s  \;&\simeq\; 1 - 6\epsilon_* + 2\eta_* \;\simeq\; 1-\frac{8(4N_*+9)}{(4N_*+3)^2}\,,\\ \label{eq:rN}
r \;&\simeq\; 16\epsilon_* \;\simeq\; \frac{192}{(4N_*+3)^2}\,.
\end{align}
The number of $e$-folds to the end of inflation are in turn related to the post-inflationary expansion history via the expression~\cite{Liddle:2003as,Martin:2010kz}
\begin{align}\label{eq:Nstar} 
N_* \;=\; &\ln\left[\frac{1}{\sqrt{3}}\left(\frac{\pi^2}{30}\right)^{1/4}\left(\frac{43}{11}\right)^{1/3}\frac{T_0}{H_0}\right]-\ln\left(\frac{k_*}{a_0H_0}\right) - \frac{1}{12}\ln g_{\rm reh} + \frac{1}{4}\ln\left(\frac{V_*^2}{\rho_{\rm end}M_P^4}\right) + \ln R_{\rm rad}\,,
\end{align}
where $H_0=67.36\,{\rm km}\, {\rm s}^{-1}{\rm Mpc}^{-1}$~\cite{Planck:2018vyg}, $T_0=2.7255\,{\rm K}$~\cite{Fixsen:2009ug} and $a_0=1$ are the present Hubble parameter, CMB temperature and scale factor, respectively. The energy density of the universe at the end of inflation is denoted by $\rho_{\rm end}$, while $g_{\rm reh}$ are the number of degrees of freedom at the end of reheating. At temperatures above the electroweak phase transition we will consider for definiteness the SM value $g_{\rm reh}=427/4$. $R_{\rm rad}$ denotes the reheating parameter~\cite{Martin:2010kz} 
\beq
R_{\rm rad} \;\equiv\;  \frac{a_{\rm end}}{a_{\rm rad}} \left(\frac{\rho_{\rm end}}{\rho_{\rm rad}}\right)^{1/4}\,,
\eeq
where $\rho_{\rm rad}$ and $a_{\rm rad}$ are the energy density and scale factor at the onset of radiation domination after the end of reheating,\footnote{For definiteness, in this work we choose $a_{\rm rad}$ such that $w(a_{\rm rad})=1/3-\delta$, where $w$ is the equation-of-state parameter of the universe, and $\delta=10^{-2}$~\cite{Ellis:2015pla}.} and $a_{\rm end}$ is the scale factor at the end of inflation. For the Starobinsky potential, substitution of (\ref{eq:phistar}), (\ref{eq:phiend}) and (\ref{eq:mphi}) results in the following non-algebraic equation,
\beq
N_* \;\simeq\; 57.76 - \frac{1}{2}\ln N_* - \left(\frac{3}{4N_*}\right)^{1+\frac{3}{4N_*}} - \ln\left(\frac{k_*}{0.05\,{\rm Mpc}^{-1}}\right) - \frac{1}{12}\ln\left(\frac{g_{\rm reh}}{427/4}\right) + \ln R_{\rm rad}\,.
\eeq
We note that for the {\em Planck} pivot scale, and SM degrees of freedom after inflation, the previous expression results in the upper bound $N_* \lesssim 55.74$, corresponding to instantaneous reheating, $R_{\rm rad}=1$. If reheating is not instantaneous, and can be modeled as the perturbative decay of the inflaton, the reheating parameter can be approximated as~\cite{Ellis:2021kad}
\beq\label{eq:Rradpert}
\ln R_{\rm rad} \;\simeq\; \frac{1}{6}\ln\left(\frac{\Gamma_{\phi}}{H_{\rm end}}\right) \;\simeq\; -0.30 + \frac{1}{3}\ln y\,, 
\eeq
where 
\beq\label{eq:gammaphi}
\Gamma_{\phi} \;=\; \frac{y^2}{8\pi}m_{\phi}\,,
\eeq
denotes the two-body inflaton decay rate, with $y$ an effective Yukawa-like coupling.

\subsection{Exact CMB observables}

An accurate determination of the inflationary observables requires the numerical computation of the primordial scalar and tensor power spectra. The necessity for this is two-fold. Firstly, Eqs.~(\ref{eq:nsN}) and (\ref{eq:rN}) are accurate only to leading order in the slow roll approximation. Secondly, and more importantly, curvature and tensor spectra do not freeze-out in value immediately upon horizon crossing, and this delay leads to a shift between the analytical and numerical relation between $(n_s,r)$ and $N_*$~\cite{Ellis:2021kad}.

Scalar fluctuations can be characterized in terms of the gauge-invariant Mukhanov-Sasaki variable, which in the Newtonian gauge is defined in terms of the inflaton fluctuation $\delta\phi$ and the scalar metric perturbation $\Psi$ as~\cite{Lalak:2007vi,Ellis:2014opa}
\beq
Q \;=\; \delta\phi + \frac{\dot{\phi}}{H}\Psi \;=\; \int \frac{d^3 \bk}{(2\pi)^{3/2}} e^{-i \bk\cdot\bx} \left[ Q_k(t) \hat{a}_{\bk} + Q_k^{*}(t) \hat{a}_{-\bk}^{\dagger} \right]
\eeq
where the dot denotes a derivative with respect to cosmic time, and $\hat{a}^{\dagger}_{\bk}$ and $\hat{a}_{\bk}$ denote creation and annihilation operators, respectively, which obey the canonical commutation relations $[\hat{a}_{\bk},\hat{a}^{\dagger}_{\bk'}]=\delta(\bk-\bk')$ and $[\hat{a}_{\bk},\hat{a}_{\bk'}]=[\hat{a}^{\dagger}_{\bk},\hat{a}^{\dagger}_{\bk'}] =0$. The mode functions $Q_k$ satisfy the equation of motion 
\beq\label{eq:MSeq}
\ddot{Q}_k + 3H\dot{Q}_k + \left[ \frac{k^2}{a^2} + \frac{3\dot{\phi}^2}{M_P^2} - \frac{\dot{\phi}^4}{2H^2M_P^4} + 2\frac{\dot{\phi}V_{\phi}}{HM_P^2}+V_{\phi\phi} \right]Q_k\;=\;0\,,
\eeq
with Bunch-Davies initial condition $Q_{k\gg aH} \;=\; e^{-ik\tau}/a\sqrt{2k}$, being $d\tau=dt/a$ the conformal time. Upon solution, the curvature perturbation is determined as
\beq
\mathcal{R} \;=\; \frac{H}{|\dot{\phi}|}Q\,,
\eeq
with power spectrum
\beq
\langle \hat{\mathcal{R}}_{\bk}\hat{\mathcal{R}}^{\dagger}_{\bk'} \rangle \;=\; \frac{2\pi^2}{k^3}\mathcal{P}_{\mathcal{R}}\,\delta(\bk-\bk')\,,
\eeq
that evaluates to
\beq
\mathcal{P}_{\mathcal{R}} \;=\; \frac{k^3 H^2}{2\pi^2 \dot{\phi}^2}|Q_k|^2\,,
\eeq
in terms of the $Q$ mode functions. The amplitude and tilt are evaluated from their definitions, not at horizon crossing, but at the end of inflation, ensuring that their freeze-out values have been reached,
\beq\label{eq:ASnsN}
A_{S*}\;=\; \mathcal{P}_{\mathcal{R}}(k_*)\big|_{a=a_{\rm end}}\,, \qquad n_s \;=\; 1+ \frac{d\ln\mathcal{P}_{\mathcal R}}{d\ln k}\bigg|_{a=a_{\rm end}}\,.
\eeq

In the case of tensor perturbations, which are automatically gauge-invariant, we use the standard transverse, traceless perturbation
\beq
h_{ij} \;=\; \int \frac{d^3\bk}{(2\pi)^{3/2}} e^{-i\bk\cdot\bx}\sum_{\gamma=+,\times} \epsilon_{ij}^{\gamma}(k)\left[ h_{\bk,\gamma}(t) \hat{b}_{\bk,\gamma} + h_{\bk,\gamma}^*(t) \hat{b}^{\dagger}_{-\bk,\gamma} \right]
\eeq
where $\gamma=+,\times$ denotes the two polarization modes, $\epsilon_{ii}^{\gamma}=k^i\epsilon_{ij}^{\gamma}=0$ and $\epsilon_{ij}^{\gamma}\epsilon_{ij}^{\gamma'}=2\delta_{\gamma\gamma'}$, and $\hat{b}_{\bk,\gamma},\hat{b}^{\dagger}_{\bk,\gamma}$ denote the creation and annihilation operators, for which $[\hat{b}_{\bk,\gamma},\hat{b}^{\dagger}_{\bk',\gamma'}]=\delta_{\gamma\gamma'}\delta(\bk-\bk')$ and $[\hat{b}_{\bk,\gamma},\hat{b}_{\bk',\gamma'}]=[\hat{b}^{\dagger}_{\bk,\gamma},\hat{b}^{\dagger}_{\bk',\gamma'}] =0$. Without anisotropic stress sources at linear order, the equation of motion satisfied by the tensor mode functions is given by
\beq\label{eq:heq}
\ddot{h}_{\bk,\gamma} + 3H\dot{h}_{\bk,\gamma} + \frac{k^2}{a^2} h_{\bk,\gamma} \;=\; 0\,.
\eeq
The corresponding power spectrum is defined as
\beq
\sum_{\gamma=+,\times} \langle \hat{h}_{\bk,\gamma}\hat{h}^{\dagger}_{\bk',\gamma}\rangle \;=\; \frac{2\pi^2}{k^3}\mathcal{P}_{\mathcal{T}}\,\delta(\bk-\bk')\,,
\eeq
from which the tensor-to-scalar ratio is computed as follows,
\beq\label{eq:rnum}
r \;=\; \left.\frac{\mathcal{P}_{\mathcal{T}}(k_*)}{\mathcal{P}_{\mathcal{R}}(k_*)}\right|_{a=a_{\rm end}}\,.
\eeq

\begin{figure}[!t]
\centering
    \includegraphics[width=\textwidth]{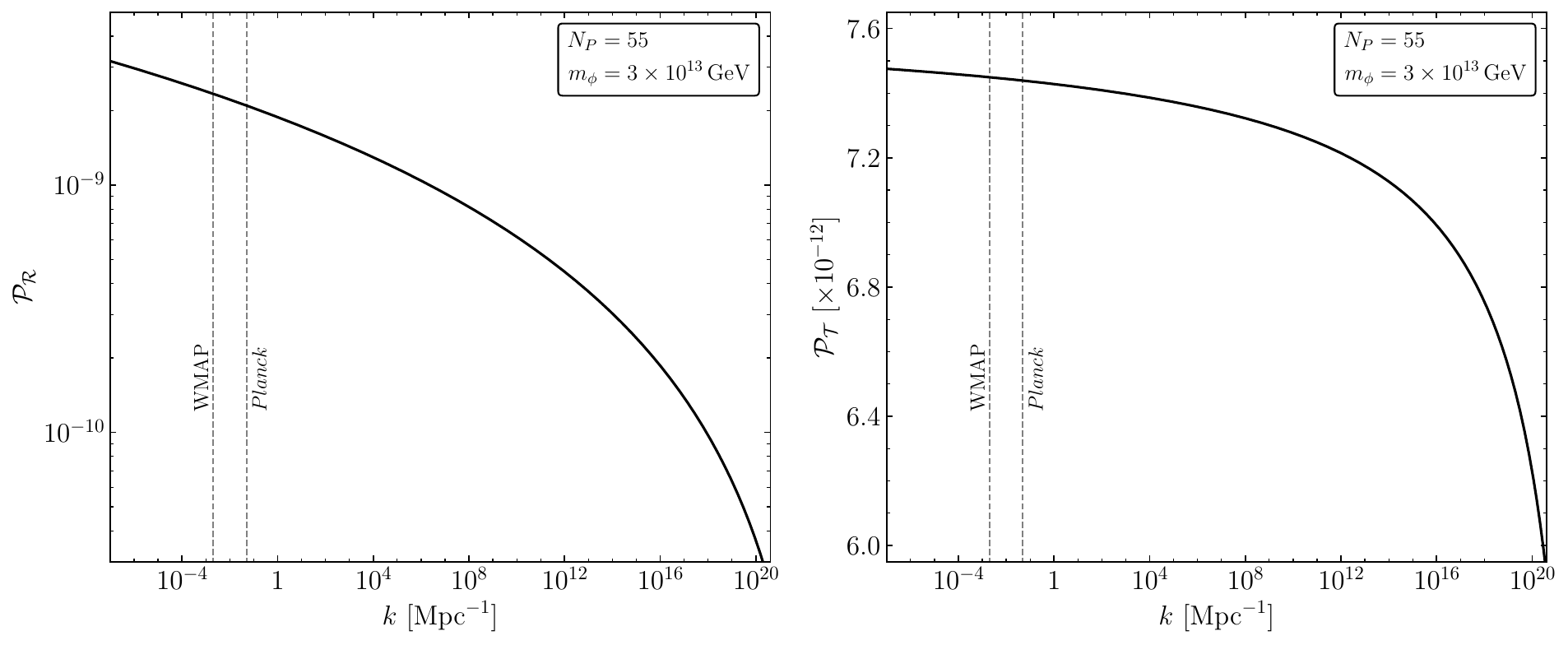}
    \caption{Numerically computed curvature (left) and tensor (right) power spectra for the Starobinsky model (\ref{eq:staropot}), assuming that the {\em Planck} pivot scale $k_P=0.05\,{\rm Mpc}^{-1}$ leaves the horizon 55 $e$-folds before the end of inflation. The inflaton mass is chosen to match the measured value of $A_{S}$. The vertical dashed lines show for comparison the location of the {\em Planck} and WMAP ($k_{\rm W}=0.002\,{\rm Mpc}^{-1}$) pivot scales.}
    \label{fig:PRPT}
\end{figure}

\begin{figure}[!t]
\centering
    \includegraphics[width=\textwidth]{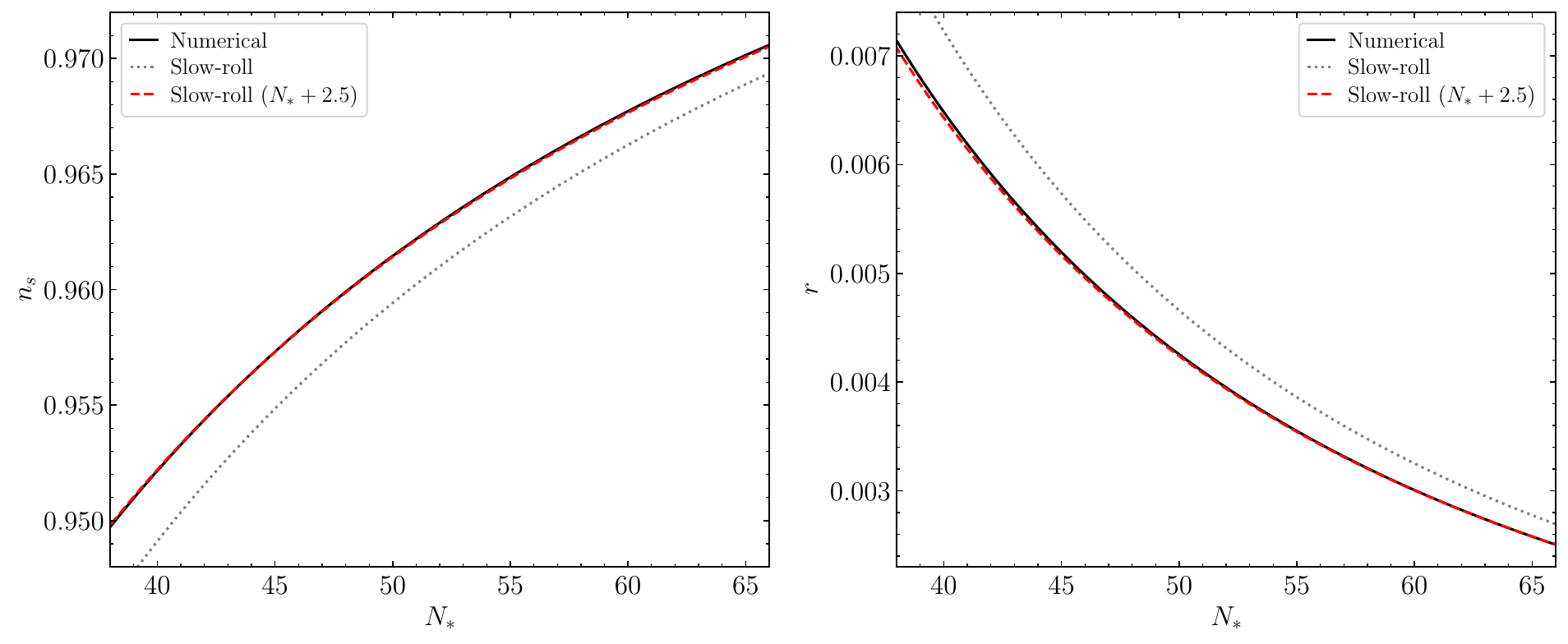}
    \caption{Scalar tilt $n_s$ (left) and tensor-to-scalar ratio $r$ (right), as functions of the number of $e$-folds after horizon crossing, $N_*$, for the Starobinsky model (\ref{eq:staropot}). The continuous black lines corresponds to the numerical evaluation of Eqs.~(\ref{eq:ASnsN}) and (\ref{eq:rnum}). The gray dotted lines are the slow-roll approximations (\ref{eq:nsN}) and (\ref{eq:rN}), evaluated at horizon crossing. The red dashed curves correspond to the slow-roll approximation, with the substitution $N_*\rightarrow N_*+2.5$.}
    \label{fig:nsr}
\end{figure}

Fig.~\ref{fig:PRPT} shows the form of the scalar and tensor power spectra for the Starobinsky model, from the numerical integration of (\ref{eq:MSeq}) and (\ref{eq:heq}). There we assume that the {\em Planck} pivot scale left the horizon 55 $e$-folds before the end of inflation. As is shown, choosing the inflaton mass as $m_{\phi}\simeq 3\times 10^{13}\,{\rm GeV}$ we recover the measured amplitude of the curvature power spectrum at $k_P$. For comparison purposes we emphasize the difference in the spectra amplitudes at the WMAP pivot scale, $k_{\rm W}=0.002\,{\rm Mpc}^{-1}$, at which the constraints on the tensor-to-scalar ratio are commonly presented~\cite{BICEP:2021xfz,Planck:2018vyg,Planck:2018jri}. Fig.~\ref{fig:nsr} shows in turn a comparison between the numerical and analytical approximations to the CMB observables $n_s,r$, here as functions of the number of $e$-folds. We note a relatively large deviation between the numerical result (black, continuous) and the slow-roll result  (\ref{eq:nsN}) and (\ref{eq:rN}) (gray, dotted). The main culprit behind this shift is the non-instantaneous freeze-out of the power spectra upon horizon crossing; the spectra continue evolving until $k/aH\ll 1$. In order to make this manifest, we also show in Fig.~\ref{fig:nsr} the red, dashed curves, which correspond to the evaluation of the slow-roll approximation with a time-delay of 2.5 $e$-folds. In this case, the agreement between the exact results, evaluated at the end of inflation, and the analytical estimates is significantly improved.

\section{Dark sector preheating}\label{sec:preheating}

We now turn to the description of the dynamics of reheating. Our discussion is based on the assumption of a perturbative decay of the inflaton into SM degrees of freedom, e.g.~through the decay of the inflaton into fermions, and the non-perturbative production of scalar dark matter. In this Section we describe the resonant production of dark matter particles, in the linear and non-linear regimes, using a combination of spectral and lattice methods. 

\subsection{Resonant production of dark matter}

Starting from the field action (\ref{eq:action}), and upon variation with respect to the DM field $\chi$, we obtain the following equation of motion,
\beq\label{eq:chieomtime}
\left( \frac{d^2}{dt^2} - \frac{\nabla^2}{a^2} + 3H\frac{d}{dt} + m_{\chi}^2 + \sigma\phi^2\right)\chi \;=\; 0\,.
\eeq
This equation can be reformulated in a quantization-friendly form by switching to conformal time and introducing the re-scaled field
\begin{align}\label{eq:Xdef}
X(\tau,\bx) \;&\equiv\; a(\tau)\,\chi(\tau,\bx)\\
\;&=\; \int \frac{d^3 \bp}{(2\pi)^{3/2}} e^{-i\bp\cdot\bx}\left[ X_p(\tau) \hat{A}_{\bp} + X^*_p(\tau)\hat{A}^{\dagger}_{-\bp}\right]\,.
\end{align}
Here $\hat{A}_{\bp}$ and $\hat{A}^{\dagger}_{\bp}$ denote the annihilation and creation operators, respectively, which satisfy the standard commutation relations $[\hat{A}_{\bp},\hat{A}^{\dagger}_{\bp'}]=\delta(\bp-\bp')$, $[\hat{A}_{\bp},\hat{A}_{\bp'}]=[\hat{A}^{\dagger}_{\bp},\hat{A}^{\dagger}_{\bp'}]=0$. The canonical commutation relations between the field operator $X$ and its conjugate momentum are in turn fulfilled if the Wronskian condition $X_pX^{*\prime}_p-X^*_pX'_p=i$ is satisfied. In terms of the $X$ mode functions, Eq.~(\ref{eq:chieomtime}) takes the form
\beq\label{eq:Xpequation}
X''_p + \omega_p^2 X_p \;=\;0 \,,
\eeq
where primes denote differentiation with respect to conformal time, and where the effective angular frequency is defined as
\beq\label{eq:omegadef}
\omega_p^2 \;\equiv\; p^2 - \frac{a''}{a} + a^2m_{\chi}^2 + \sigma a^2 \phi^2\,. 
\eeq
The initial conditions for the $X_p$ are taken to be the Bunch-Davies vacuum mode functions,
\beq
X_p(\tau_0) \;=\; \frac{1}{\sqrt{2\omega_p}}\,,\qquad X_p'(\tau_0) \;=\; -\frac{i\omega_p}{\sqrt{2\omega_p}}\,.
\eeq
At any later time, the UV-regular number and energy densities of the field $\chi$ can be computed as follows~\cite{Kofman:1997yn,Garcia:2021iag}
\begin{align}
n_{\chi} \;&=\; \int \frac{d^3\bp}{(2\pi)^3a^3}\, f_{\chi}(p)\,,\\ \label{eq:rhochidef}
\rho_{\chi} \;&=\; \int \frac{d^3\bp}{(2\pi)^3a^4}\, \omega_p f_{\chi}(p)\,,
\end{align}
where
\beq\label{eq:fchi}
f_{\chi}(p) \;=\; \frac{1}{2\omega_p}\left| \omega_pX_p - i X'_p \right|^2\,,
\eeq
denotes the phase space distribution (PSD) of the field $\chi$, identical to the particle occupation number of the DM.

The effective frequency (\ref{eq:omegadef}) can be equivalently rewritten in terms of the instantaneous Hubble parameter as $\omega_p^2=p^2 + a^2m_{\rm eff}^2$, where
\beq\label{eq:meff}
m_{\rm eff}^2 \;\equiv\; m_{\chi}^2 + \sigma\phi^2 - \left(\frac{1-3w}{2}\right)H^2\,, 
\eeq
and $w=P/\rho$ denotes the total equation-of-state parameter~\cite{Garcia:2023awt}. For sufficiently small bare mass $m_{\chi}$, coupling $\sigma$, and momentum $p$, the last term in (\ref{eq:meff}) might dominate during inflation ($w\simeq -1$), leading to the tachyonic excitation of super-horizon modes. Assuming a DM mass smaller than the Hubble parameter during inflation, $m_{\chi}\ll H$, the tachyonic instability would occur if 
\beq
p^2 \;\lesssim\; a^2\left( \frac{1-3w}{2}H^2 - \sigma \phi^2\right)\,.
\eeq
In other words, for 
\beq\label{eq:tachyoniccond}
\frac{\sigma}{\lambda} \;\gtrsim\; \frac{1-3w}{2}\left(\frac{H^2}{\lambda M_P^2}\right)\left(\frac{M_P}{\phi}\right)^2\,,
\eeq
where 
\beq
\lambda \;\equiv\; m_{\phi}^2/2M_P^2\,, 
\eeq
the non-perturbative growth of low-momentum modes during inflation can be safely ignored. For the Starobinsky model, evaluation of (\ref{eq:tachyoniccond}) at the end of inflation leads to $\sigma/\lambda\gtrsim 0.5$. In what follows, we will assume that the effective coupling $\sigma/\lambda \gg 1$, and therefore we will disregard the evolution of the $\chi$ mode functions prior to the beginning of inflation. 

In the large coupling regime, the dominant term in the effective mass corresponds to $\sigma\phi^2$, which is quasi-periodic in time, leading to the parametric excitation of momentum modes. The solution to the equation of motion (\ref{eq:KGeq}) near the minimum of the potential is approximately given by
\beq
\phi(t) \;\simeq\; \phi_0(t)\cdot \cos(m_{\phi}t)\,,
\eeq
with $\phi_0\simeq \frac{\phi_{\rm end}}{m_{\phi}t}$ denoting a decreasing envelope. It is then convenient to introduce the following variables,
\begin{alignat}{2}
  x_q(t,\bx) & \;=\; a(t)^{1/2}X_q(t,\bx)\,,
  & \qquad\qquad \kappa & \;=\; \frac{1}{8}\left(\frac{\sigma}{\lambda}\right)\left(\frac{\phi_0}{M_P}\right)^2\,, \\
  z & \;=\; m_{\phi}t+{ \frac{\pi}{2}}\,,
  & A_{q} & \;=\; q^2\left(\frac{a_{\rm end}}{a}\right)^2 + 2\kappa\,,
\end{alignat}
where
\beq
q \;=\; \frac{p}{m_{\phi}a_{\rm end}}\,,
\eeq
denotes the dimensionless comoving momentum relative to the end of inflation, rescaled by with respect to the inflaton mass. Eq.~(\ref{eq:Xpequation}) takes then the Mathieu form
\beq\label{eq:Mathieu}
\frac{d^2 x_q}{dz^2} + (A_{q} - 2\kappa \cos 2z)x_q \;=\; 0\,,
\eeq
which presents parametric resonance. More precisely, Floquet's theorem states that the solutions of this equation have the form~\cite{Magnus2004-br}
\beq
{ x_q(z) \;=\; e^{\mu_{q}z}g_1(z) + e^{-\mu_{q}z}g_2(z)}\,.
\eeq
\begin{figure}[!t]
    \includegraphics[width=0.35\textwidth]{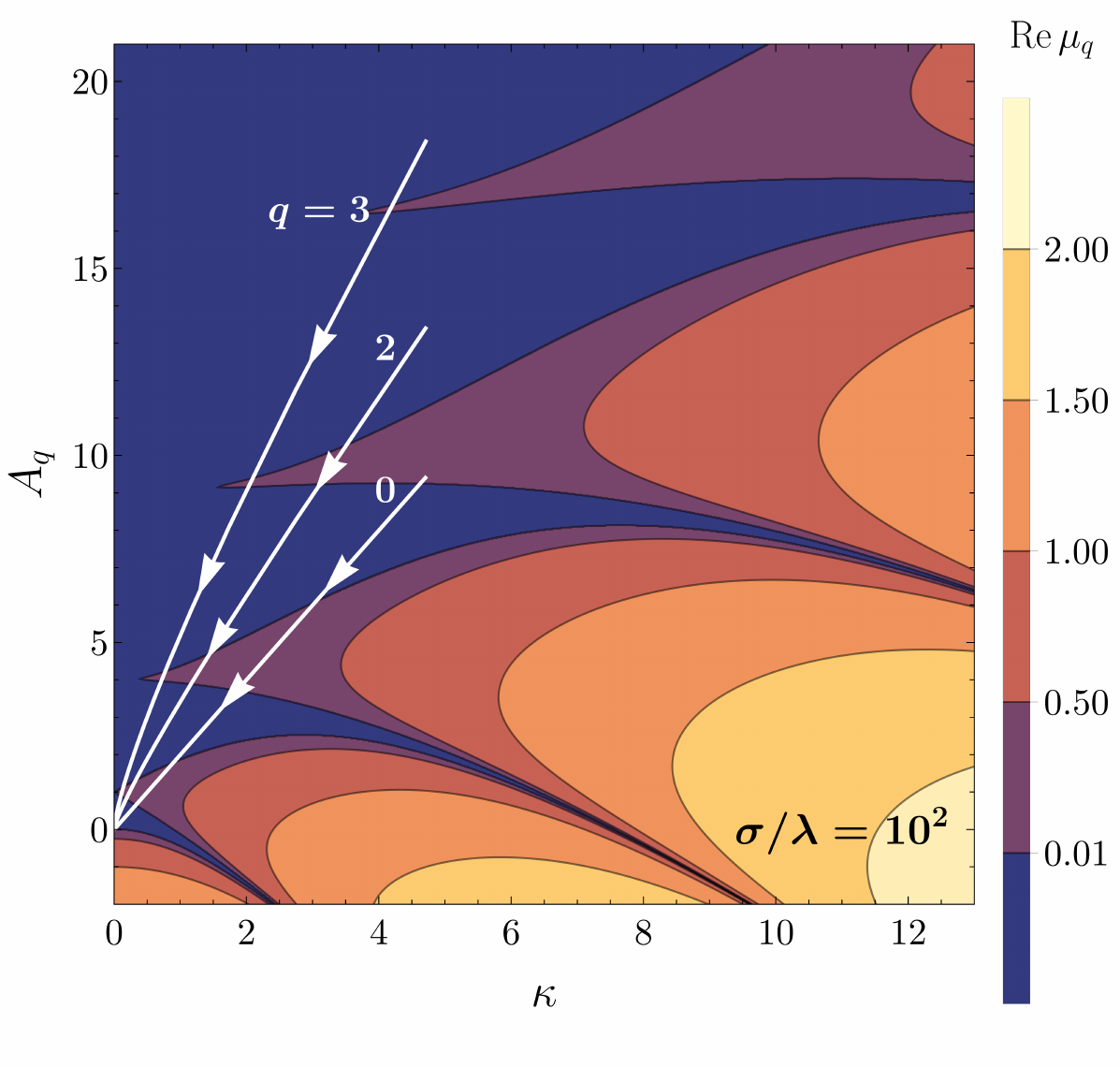}
    \includegraphics[width=0.65\textwidth]{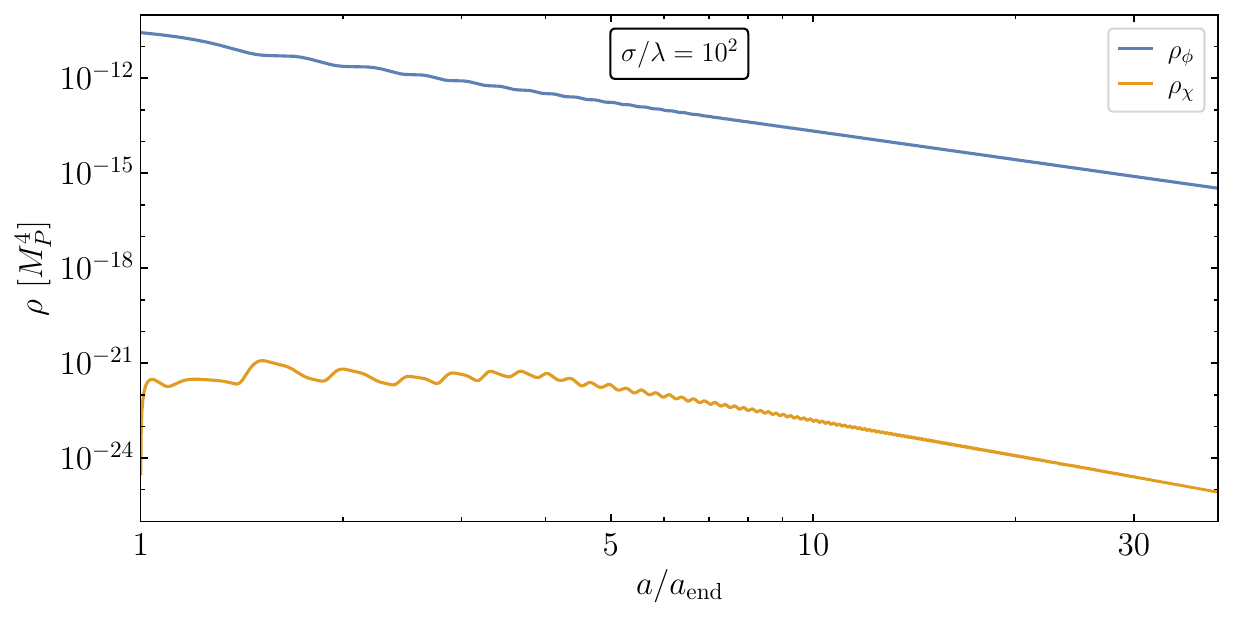}\\
    \includegraphics[width=0.35\textwidth]{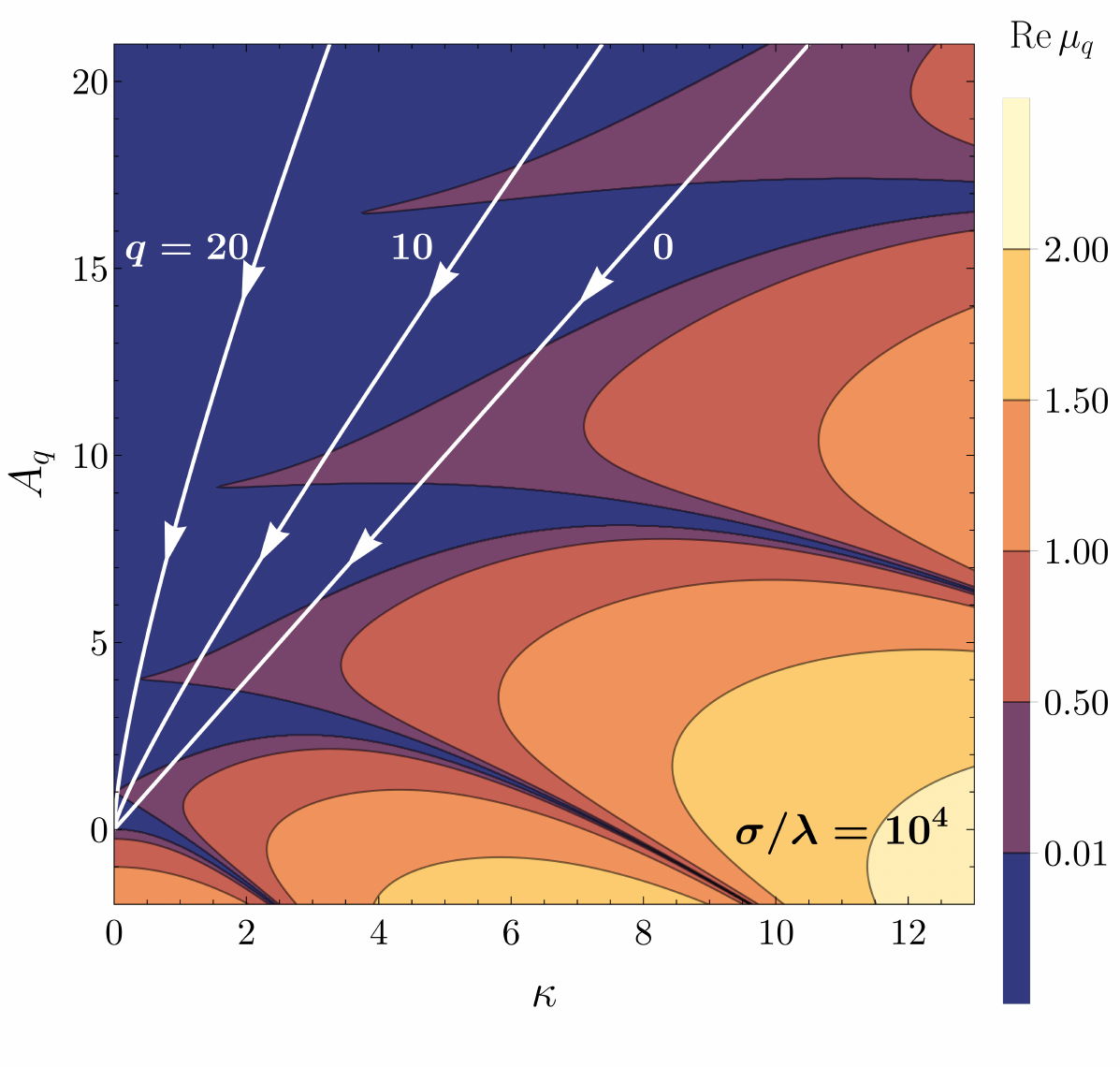}
    \includegraphics[width=0.65\textwidth]{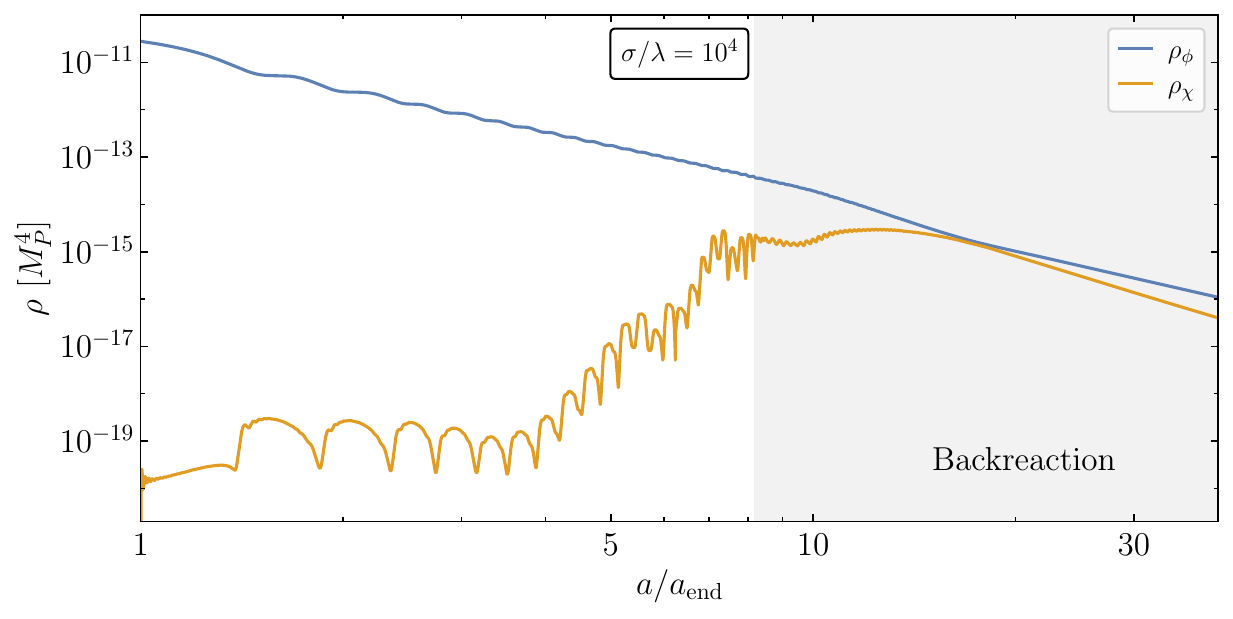}
    \caption{Floquet charts for the Mathieu equation (\ref{eq:Mathieu}) showing the flow lines for selected comoving momenta for the Starobinsky potential (\ref{eq:staropot}) (left), together with the corresponding evolution of the energy densities of the inflaton and DM during reheating (right), in the absence of backreaction (top) and in the presence of backreaction and fragmentation (bottom).}
    \label{fig:floquet}
\end{figure}
Here $g_1(z)$ and $g_2(z)$ are periodic functions, and $\mu_q$ is called the Floquet exponent. If ${\rm Re}\,\mu_q \;\neq 0$ exponentially growing solutions exist. The Floquet exponents, and the corresponding resonance bands as a function of $A_q$ and $\kappa$ can be found following the eigenvalue method described in detail in~\cite{Amin:2014eta}. The left panels of Fig.~\ref{fig:floquet} show the structure of the resonance bands of the Mathieu equation (\ref{eq:Mathieu}), with a color coded Floquet exponent. Since the parameters $A_q$ and $\kappa$ depend on redshifting quantities, each momentum mode flows on this chart, from an initial point determined by $\phi_{\rm end}$ and the effective coupling $\sigma/\lambda$, to the origin. The figure shows the corresponding flows for a few selected (re-scaled) momenta $q$ for $\sigma/\lambda=10^{2},10^4$. For the first case, low-momentum modes only cross a couple of resonance bands, which is manifest in the behavior of the energy density of the DM field, shown in the right panel. The resonant growth of $\rho_{\chi}$ is not accumulated over several oscillations, and the $\chi$ field enters the pure redshift regime, with $\rho_{\chi}\propto a^{-4}$, for $a/a_{\rm end}\gtrsim 4$. On the other hand, for $\sigma/\lambda=10^4$, each low momentum mode can cross a large number of resonance bands, and the resonance can efficiently accumulate. As shown in the right panel, this results in the quasi-exponential growth of the energy density of the DM, which becomes quickly comparable with $\rho_{\phi}$. The linear description in terms of Eq.~(\ref{eq:Xpequation}) breaks down and the system then enters the backreaction regime, which we describe below. 

\subsection{Backreaction and fragmentation}

As demonstrated above in Fig.~\ref{fig:floquet}, for large effective couplings $\sigma/\lambda \gg 1$, the resonant growth of the $\chi$ mode functions can take them into the backreaction regime. The energy density of $\chi$ becomes comparable to the energy density of the inflaton, and the DM can no longer be treated as an spectator field. More importantly, the explosive production of DM can disrupt the homogeneous inflaton condensate by means of the rescattering of $\chi$ particles into $\phi$, i.e.~the production of inflaton particles. Moreover, the free DM particles can also scatter inflaton fluctuations away from the condensate. At the level of fields, these dynamics are manifested as the growth of spatial gradients of the inflaton field, eventually leading to the ``fragmentation'' of the homogeneous condensate~\cite{Garcia-Bellido:2002fsq,Felder:2006cc,Frolov:2010sz,Amin:2014eta,Garcia:2021iag}.

In terms of the field mode functions, these effects manifest as mode-mode couplings, since now the inflaton field in (\ref{eq:chieomtime}) is to be replaced by $\phi(t)\rightarrow \phi(t) + \delta\phi(t,\bx)$, where the fluctuation $\delta\phi$ is also quantized. This of course leads to a set of non-linear Heisenberg equations of motion for the $\phi,\chi$ quantum fields, which is challenging to solve. For this reason, spectral methods are not the tool of choice in the backreaction regime. Instead, we resort to the solution of the classical equations of motion for the coupled fields. At large couplings the occupation numbers for the resonantly excited momentum modes are $f_{\chi,\phi}\gg 1$, justifying this classical approach. The system of non-linear partial differential equations
\begin{align} \label{eq:lattice1}
\ddot{\phi} + 3H\dot{\phi} - \frac{\nabla^2\phi}{a^2} + \partial_{\phi}V(\phi,\chi) \;&=\; 0\,,\\
\ddot{\chi} + 3H\dot{\chi} - \frac{\nabla^2\chi}{a^2} + \partial_{\chi}V(\phi,\chi) \;&=\; 0\,,\\
\rho_{\phi} + \rho_{\chi} \;&=\; 3H^2M_P^2\,,
\end{align}
with 
\begin{align} \label{eq:lattice4}
\rho_{\phi} + \rho_{\chi} \;&=\; \frac{1}{2}\dot{\phi}^2 + \frac{1}{2a^2}(\nabla \phi)^2 + \frac{1}{2}\dot{\chi}^2 + \frac{1}{2a^2}(\nabla \chi)^2 + V(\phi,\chi)\,,
\end{align}
is solved by means of finite-difference techniques on a spatial lattice. Among the variety of publicly available codes~\cite{Felder:2000hq,Frolov:2008hy,Sainio:2012mw,Lozanov:2019jff}, we favor \CL~\cite{Figueroa:2020rrl,Figueroa:2021yhd} due to its ease of use. The lower right panel of Fig.~\ref{fig:floquet} shows precisely the lattice solution of the system (\ref{eq:lattice1})-(\ref{eq:lattice4}) for $\sigma/\lambda=10^4$. For definiteness we have associated the presence of backreaction with the condition $\rho_{\chi}\gtrsim 0.1\,\rho_{\phi}$~\cite{Garcia:2021iag,Garcia:2022vwm}. Since this system is not dissipative, the inflaton and the DM exchange energy when the backreaction regime is reached, entering a state of quasi-equilibrium until the expansion of the universe overcomes the rescattering rates, and the system enters the pure redshift regime, with $\rho_{\phi}\propto a^{-3}$ and $\rho_{\chi}\propto a^{-4}$. 

\subsection{Distributions and densities}

\begin{figure}[!t]
    \includegraphics[width=\textwidth]{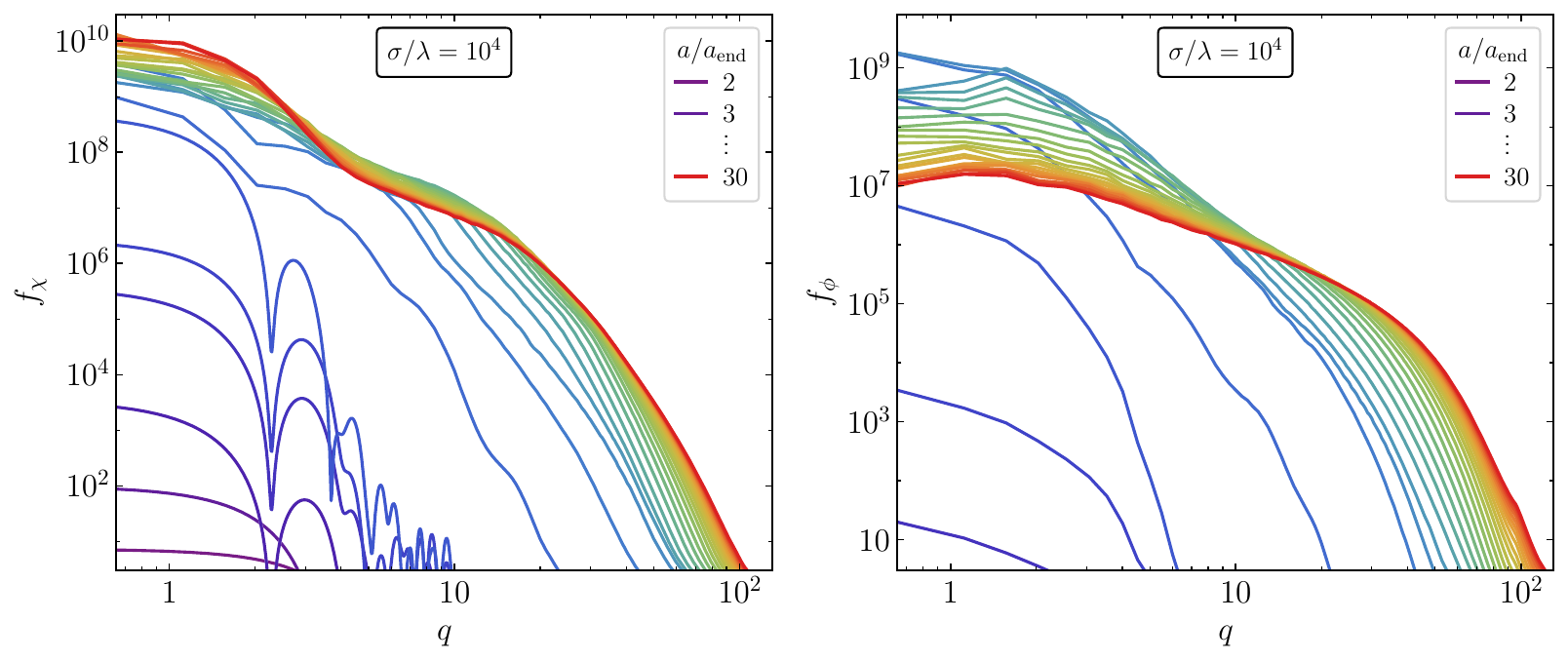}
    \includegraphics[width=\textwidth]{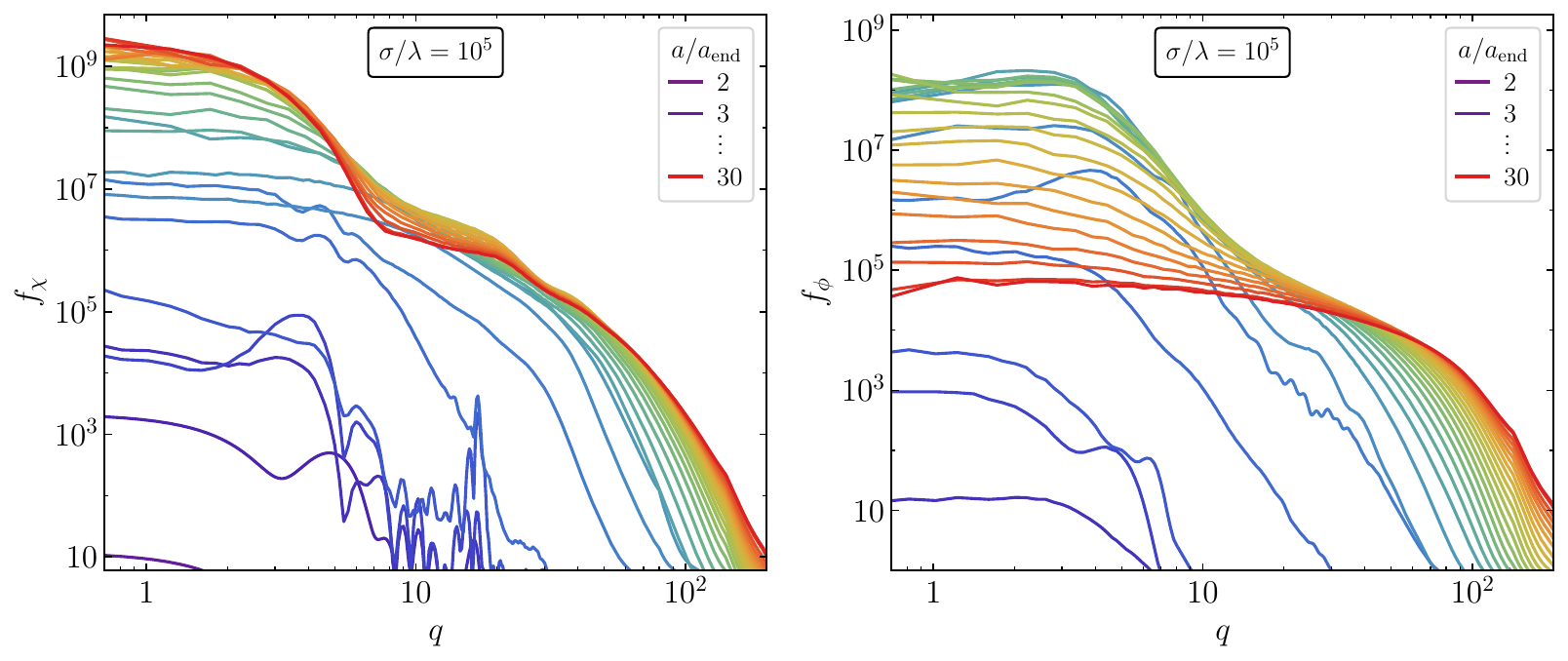}
    \caption{Phase space distributions for the DM (left) and inflaton (right) fields, as functions of the comoving momentum $q$, evaluated at particular values of $a/a_{\rm end}$. Top: PSDs for $\sigma/\lambda=10^4$, as obtained from \CL~with $512^3$ lattice points and $k_{\rm IR}=0.5$. Bottom: PSDs for $\sigma/\lambda=10^5$ with $800^3$ lattice points and $k_{\rm IR}=0.2$.}
    \label{fig:psd1}
\end{figure}

\begin{figure}[!t]
    \includegraphics[width=\textwidth]{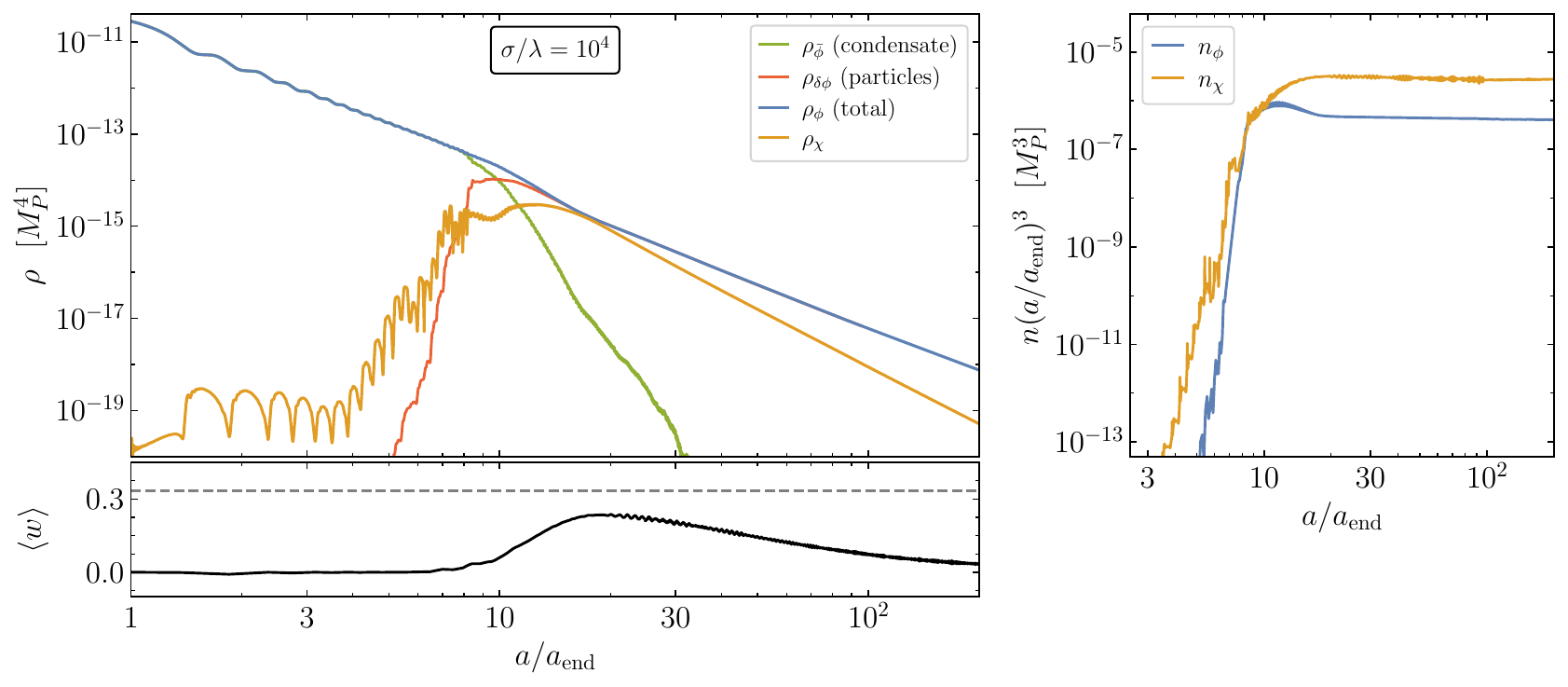}
    \caption{Top, left: Evolution of the total inflaton energy density (blue), the energy density of the inhomogeneous inflaton component (red), the energy density of the homogeneous inflaton component (green), and the DM energy density (yellow), for $\sigma/\lambda=10^4$. Right: Comoving number densities for the inflaton quanta (blue) and the DM quanta (yellow). Bottom, left: the oscillation-averaged total equation-of-state parameter.}
    \label{fig:rhon1}
\end{figure}

\begin{figure}[!t]
    \includegraphics[width=\textwidth]{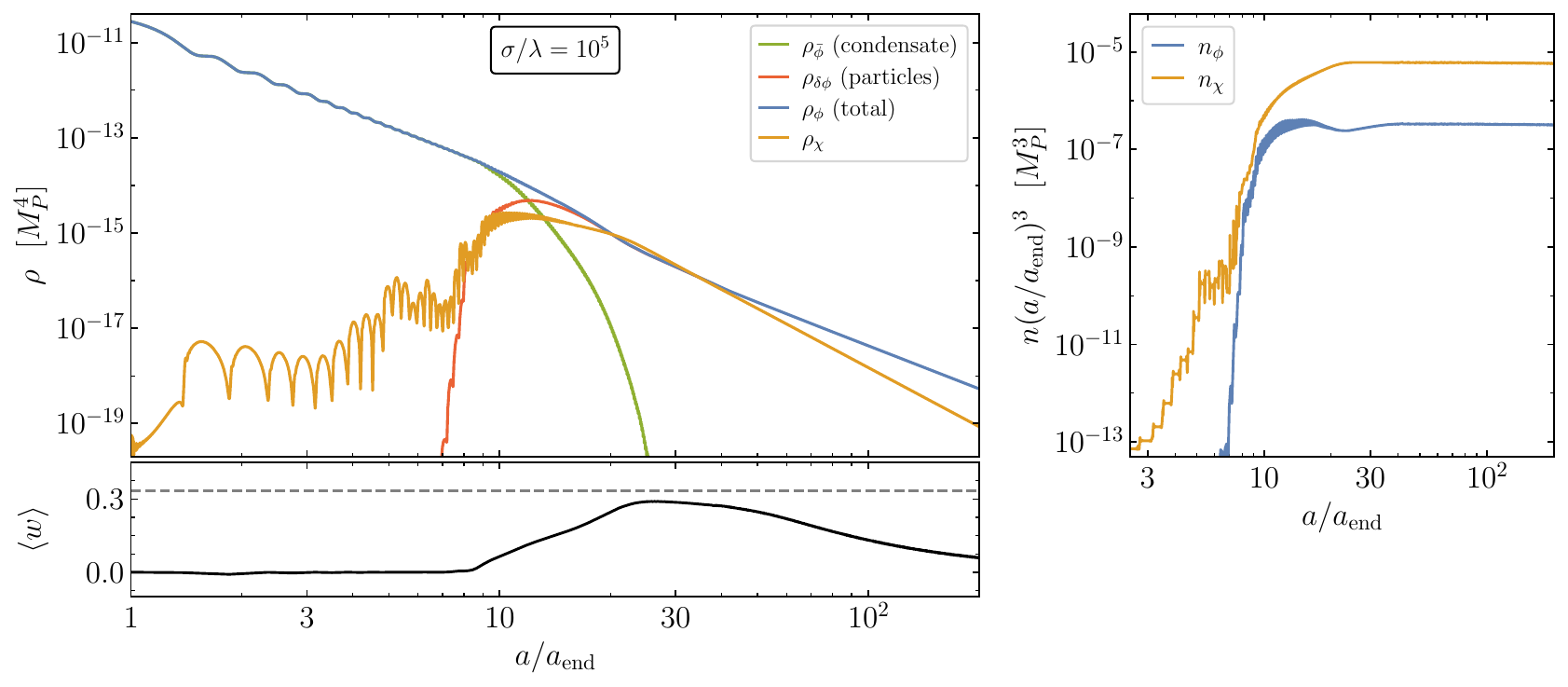}
    \caption{Same as Fig.~\ref{fig:rhon1}, for $\sigma/\lambda=10^5$.}
    \label{fig:rhon2}
\end{figure}

By means of a combination of spectral and lattice methods we can track the time evolution of energy and number densities of fields, and of their corresponding PSDs. The top panels of Fig.~\ref{fig:psd1} show the PSDs of the DM and the inflaton as functions of the scale factor, for the coupling $\sigma/\lambda=10^4$ (and $\sigma/\lambda=10^5$ for the bottom panels).\footnote{ The number of lattice points has been chosen from dedicated convergence tests, to ensure independence of the PSDs, and number and energy densities, from the infrared and ultraviolet momentum cutoffs induced by the lattice size and spacing.} Focusing on $\chi$, at early times, $a/a_{\rm end}\lesssim 9$ ($a/a_{\rm end}\lesssim 5$), the growth of the distribution follows the spectral result as given by (\ref{eq:fchi}). The distribution clearly shows a peak pattern produced by the resonant growth owing to the coupling of $\chi$ to the oscillating inflaton. This growth ``in bursts'' can also be appreciated in the DM energy and number densities, which are displayed in Figs.~\ref{fig:rhon1} and \ref{fig:rhon2}. At later times, the backreaction regime is reached, and the efficient energy exchange between the dark and inflaton sectors rapidly erases the resonance peaks of the distribution. The distribution quickly evolves towards the UV, with two plateau-like regions owing to Kolmogorov-like turbulence~\cite{Micha:2002ey,Micha:2004bv}, and a quasi-thermal exponential tail. Relativistic modes come to dominate the energy density of $\chi$, and eventually the pure redshift regime $\rho_{\chi}\propto a^{-4}$ is reached. This signals the end of the backreaction regime, and the return to dominance of $\phi$. The PSD freezes, and so does the comoving number density, as shown in the right panels of Figs.~\ref{fig:rhon1} and \ref{fig:rhon2}. In the absence of DM interactions other than its coupling to $\phi$, the number density produced during reheating is only redshifted to become the present DM relic abundance. We postpone this discussion to Section~\ref{sec:DM}.

Focusing now on the inflaton $\phi$, the right panels of Fig.~\ref{fig:psd1} show how the production of non-zero modes of $\phi$ (i.e.~particles) is induced by rescattering. Even before backreaction is reached we have $f_{\phi}\gg 1$ for low momentum modes. Initially, the production of non-relativistic modes is most efficient, until a maximum in the PSD is reached when backreaction is reached. Posterior to this, the cascading of energy toward the UV is efficiently mediated by rescattering, until the PSD freezes when $\rho_{\chi}\ll \rho_{\phi}$. The comoving number density of inflaton quanta remains also frozen, as it can be observed in the right panels of Figs.~\ref{fig:rhon1} and \ref{fig:rhon2}. The depletion of the inflaton zero-mode in favor of $q\neq0$ modes can be more explicitly visualized in the upper left panels of Figs.~\ref{fig:rhon1} and \ref{fig:rhon2}. In them, three energy densities for the inflaton have been identified. The total energy density $\rho_{\phi}$ is computed from the {\em spatial} average of the full energy momentum tensor of $\phi$,
\beq
\rho_{\phi} \;=\; \overline{\frac{1}{2}\dot{\phi}^2 + \frac{1}{2a^2}(\nabla \phi)^2 + V(\phi)}\,.
\eeq
On the other hand, we identify the condensate component of the energy density of $\phi$ as the energy density of the spatially averaged field~\cite{Garcia:2021iag,Garcia:2023eol,Garcia:2023dyf},
\beq
\rho_{\bar{\phi}} \;=\; \frac{1}{2}\dot{\bar{\phi}}^2 + V(\bar{\phi})\,.
\eeq
The energy density of inflaton fluctuations can then simply be defined as 
\beq
\rho_{\delta\phi} \;=\; \rho_{\phi}-\rho_{\bar{\phi}}\,,
\eeq
whose value is consistent at early times with the spectral definition of Eq.~(\ref{eq:rhochidef}), replacing $\chi$ by $\phi$~\cite{Garcia:2023eol}. In terms of these ``components'' of $\rho_{\phi}$ it is easy to see the onset of backreaction in the oscillating condensate. Non-zero modes are efficiently populated by rescatterings, draining energy density from $\bar{\phi}$ until $\rho_{\delta\phi}> \rho_{\bar{\phi}}$. At this point, the inflaton can be considered as fragmented in favor of free quanta. As the bottom panels of Figs.~\ref{fig:rhon1} and \ref{fig:rhon2} also show, the rapid population of relativistic $\phi$ and $\chi$ modes drives the oscillation-averaged equation-of-state parameter $\langle w\rangle$ close to its radiation-domination value. Importantly, the depletion of $\rho_{\bar{\phi}}$ is rapid and complete. After fragmentation the inflaton is fully replaced by free particles, which eventually are redshifted by expansion and, in the absence of other interactions, lead again to a matter dominated era, with $\rho_{\phi}\simeq\rho_{\delta\phi}\simeq m_{\phi}n_{\phi}$. 

The end of backreaction, and the subsequent dominance of nonrelativistic free inflaton quanta, indicate that reheating cannot be completed by dark sector preheating. The couplings contained in $\mathcal{L}_{\rm SM}$ in (\ref{eq:action}) are crucial to fully dissipate $\phi$ and reheat the universe. In the next Section we study the decay of the fragmented inflaton, and the effect that the nonlinear dynamics during preheating have on the duration of reheating, the dark matter abundance and the inflationary CMB observables.

\section{Signatures of preheating}\label{sec:signatures}

Having characterized the effect of dark sector preheating on the evolution of the background metric and inflaton energy densities, we now proceed to quantify its effect on the reheating process, understood here as the dissipation of the inflaton energy density into SM relativistic degrees of freedom. We begin by reviewing the Boltzmann formalism for perturbative reheating, to continue with the determination of DM relic abundances, and the effects of preheating in CMB observables. 

\subsection{Reheating after inflaton fragmentation}\label{sec:reheatingfrag}

As we briefly mentioned in the Introduction, we will assume that the decay of the inflaton into the visible sector is driven by the perturbative decay of the inflaton quanta. For simplicity, we will model this process as a two-body decay into a pair of fermions $f$, with effective Yukawa coupling $y$,
\beq
\mathcal{L}_{\rm SM} \;\supset\; y\phi\bar{f}{f}\,.
\eeq
If the mass of the final state particles can be neglected, and assuming that quantum statistics can be disregarded, the Boltzmann equation that determines the population of the primordial radiation plasma with energy density $\rho_R$ is given by~\cite{Garcia:2023eol}
\beq\label{eq:boltzR}
\dot{\rho}_R + 4H\rho_R \;=\; \Gamma_{\phi}(\rho_{\bar{\phi}} + m_{\phi}n_{\delta\phi})\,,
\eeq
where $\Gamma_{\phi}$ is the vacuum inflaton decay rate (\ref{eq:gammaphi}). The first term in the right hand side corresponds to the contribution from the dissipation of the coherent inflaton condensate, with $w_{\phi}=0$. The second term represents the decay of the free particles that compose the fragmented inflaton after backreaction. It is worth recalling here that prior to fragmentation (or in the absence of it), $\rho_{\phi}\simeq \rho_{\bar{\phi}}$, while after fragmentation, $\rho_{\phi}\simeq m_{\phi}n_{\delta\phi}$, as shown in Fig.~\ref{fig:rhon1}.

Disregarding the coupling of $\phi$ to $\chi$, which is valid well before or well after the backreaction regime with $w\simeq0$, Eq.~(\ref{eq:boltzR}) is complemented by the system of equations
\begin{align} \label{eq:BoltzR}
\dot{\rho}_{\phi} + 3H\rho_{\phi} \;&=\; -\Gamma_{\phi}(\rho_{\bar{\phi}} + m_{\phi}n_{\delta\phi})\,,\\ \label{eq:boltzH}
3H^2M_P^2 \;&=\; \rho_{\phi} + \rho_R\,,
\end{align}
from which the dynamics of the background quantities can be fully determined. The end of reheating is defined as the crossover time $t_{\rm reh}$, when $\rho_{\phi}(t_{\rm reh})=\rho_R(t_{\rm reh})$. In turn, the reheating temperature $T_{\rm reh}$ is defined from the thermodynamic relation
\beq
\rho_R \;=\; \frac{\pi^2}{30}gT^4\,,
\eeq
evaluated at $t=t_{\rm reh}$. In the absence of dark sector preheating, Eqs.~(\ref{eq:boltzR})-(\ref{eq:boltzH}) can be solved approximately during reheating. For $t\ll t_{\rm reh}$, $\rho_{\phi}\simeq \rho_{\rm end}(a_{\rm end}/a)^3$, and (\ref{eq:boltzR}) can be rewritten as
\beq
\frac{d}{da}\left(\rho_R a^4\right) \;=\; \frac{\Gamma_{\phi}\rho_{\phi}a^3}{H} \;\simeq\; \sqrt{3M_P^2\Gamma_{\phi}^2\rho_{\phi}}\,a^3\,,
\eeq
with solution~\cite{Garcia:2020eof}
\beq
\rho_{R}(a) \;\simeq\; \frac{2}{5}\sqrt{3M_P^2\Gamma_{\phi}^2\rho_{\rm end}}\left(\frac{a_{\rm end}}{a}\right)^{4}\left[\left(\frac{a}{a_{\rm end}}\right)^{5/2}-1\right]\,.
\eeq
This expression implies that the maximum temperature of the primordial plasma, assuming instantaneous thermalization, is given by
\beq \label{eq:Tmax}
T_{\rm max}\;\simeq\; \frac{5}{3}\left(\frac{3}{8}\right)^{8/5}\sqrt{3M_P^2\Gamma_{\phi}^2\rho_{\rm end}} \,, \quad{\rm at}\quad \frac{a_{\rm max}}{a_{\rm end}}\;\simeq\; \left(\frac{8}{3}\right)^{2/5}\,.
\eeq
In turn, the reheating temperature can be approximated by
\beq\label{eq:TrehP}
T_{\rm reh} \;\simeq\; \left(\frac{72 \Gamma_{\phi}^2 M_P^2}{5\pi^2 g_{\rm reh}}\right)^{1/4} \;\simeq\; y\left(\frac{9 m_{\phi}^2 M_P^2}{40 \pi^4 g_{\rm reh}}\right)^{1/4} \,, \quad{\rm at}\quad \left(\frac{a_{\rm reh}}{a_{\rm end}}\right)_{\rm pert} \;\simeq\; \left(\frac{25}{12}\frac{\rho_{\rm end}}{\Gamma_{\phi}^2M_P^2}\right)^{1/3}\,.
\eeq
As is well know, $T_{\rm reh}$ depends only on the couplings and number of degrees of freedom of the theory if $a_{\rm reh}\gg a_{\rm end}$. Therefore, we do not expect this temperature to be dependent on the details of preheating, as long as the complete decay of the inflaton does not occur during fragmentation.\par\medskip

\begin{figure}[!t]
\centering
    \includegraphics[width=0.95\textwidth]{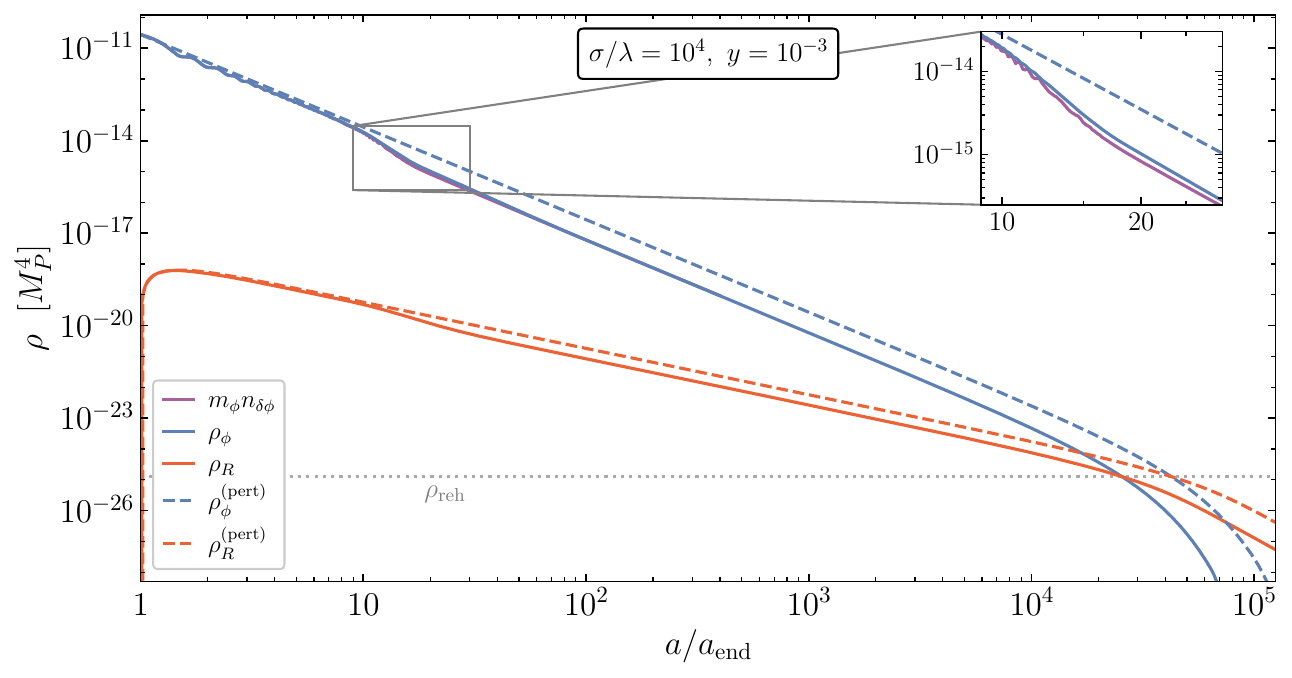}
    \caption{Inflaton and radiation energy densities after the end of inflation, in the presence of dark sector preheating (continuous) and in its absence (dashed). For the continuous curves, the full inflaton energy density is shown as the blue line, while the contribution of non-relativistic modes is shown in purple. Their difference is highlighted in the inset. The horizontal dotted line corresponds to the energy density at the end of reheating, $\rho_{\rm reh}=\rho_{\phi}(a_{\rm reh})=\rho_R(a_{\rm reh})$. }
    \label{fig:reheating}
\end{figure}

In the presence of dark sector preheating, the predictions for the duration and final temperature of reheating are modified, as is shown in Fig.~\ref{fig:reheating} for $\sigma/\lambda=10^4$. The perturbative approximation discussed above is shown as the dashed blue and red lines, labeled as $\rho_{\phi}^{({\rm pert})}$ and $\rho_R^{({\rm pert})}$, respectively. On the other hand, the energy densities of the inflaton and the radiation in the presence of non-perturbative DM production are shown as the continuous blue and red curves, respectively. We observe that, prior to the onset of backreaction, for $a/a_{\rm end}\lesssim 8$, the exact energy densities follow closely the perturbative approximation, with a small deviation originating from the transition from the quasi-de Sitter epoch to matter domination within the first oscillation of $\phi$. Notably, the maximum of $\rho_R$ is essentially identical in the two cases, and therefore the approximation (\ref{eq:Tmax}) remains valid in the presence of a strong $\phi$-$\chi$ coupling. 

As it can be appreciated in Fig.~\ref{fig:reheating}, and also Fig.~\ref{fig:rhon1}, $\chi$ backreaction is important for $8\lesssim a/a_{\rm end}\lesssim 100$. During this time, a significant fraction of the energy density of the inflaton condensate is transferred to DM and free inflaton particles. For $a/a_{\rm end}\gtrsim 10$, $\rho_{\delta\phi}>\rho_{\bar{\phi}}$, and inflaton particles make up almost the entirety of $\rho_{\phi}$. Nevertheless, only a fraction of them are non-relativistic, due to the UV-cascading induced by mode-mode couplings (see Fig.~\ref{fig:psd1}). The result is the hierarchy $\rho_{\phi}^{({\rm pert})}>\rho_{\phi}>m_{\phi}n_{\delta\phi}\gg\rho_{\bar{\phi}}$, which suppresses the effective particle production rate, as per~(\ref{eq:BoltzR}). Therefore, a reduction in the radiation energy density is expected, and it can be appreciated in the continuous red curve. In this regime, the time-dependence of all background quantities must be determined numerically, from the lattice results and the continuity equation (\ref{eq:boltzR}).

Backreaction ends when rescattering is strongly suppressed by the expansion of the universe. For $\sigma/\lambda=10^4$, this corresponds to $a/a_{\rm end}\gtrsim 100$, as shown in Fig.~\ref{fig:reheating}. In this regime, $\rho_{\phi}\simeq \rho_{\delta\phi}\simeq m_{\phi}n_{\delta\phi}$. The energy density of DM is no longer comparable to $\rho_{\phi}$, and the main particle production process is now the perturbative decay into radiation. Therefore, the system (\ref{eq:boltzR})-(\ref{eq:boltzH}) with $\rho_{\bar{\phi}}\simeq0$ accurately describes the evolution of the inflaton-radiation mixture even beyond the end of reheating. Let us denote by 
\beq
\Delta \;=\; \frac{\rho_{\phi}(a_m)a_m^3}{\rho_{\rm end} a_{\rm end}^3} \;\simeq\; \begin{cases}
0.22\,,\ & \sigma/\lambda=10^4\,,\\
0.15\,,\ & \sigma/\lambda=10^5\,,\\
0.07\,,\ & \sigma/\lambda=10^6\,,
\end{cases}
\eeq
the loss in the comoving inflaton energy density between the end of inflation ($a_{\rm end}$) and the end of backreaction ($a_m$), due to $\chi$-preheating. Then, we can modify (\ref{eq:TrehP}) to estimate the change in the reheating temperature and the duration of reheating due to the depletion in $\rho_{\phi}$. We note that the reheating temperature, being independent of $\rho_{\rm end}$ is insensitive to preheating, while the duration is modified to
\beq\label{eq:arehpreh}
\left(\frac{a_{\rm reh}}{a_{\rm end}}\right)_{\rm preh} \;\simeq\; \left(\frac{25}{12}\frac{\Delta\,\rho_{\rm end}}{\Gamma_{\phi}^2M_P^2}\right)^{1/3} \;=\; \Delta^{1/3}\left(\frac{a_{\rm reh}}{a_{\rm end}}\right)_{\rm pert}\,.
\eeq
For the parameters in Fig.~\ref{fig:reheating}, our analytical estimates give $\rho_{\rm reh}\simeq 1.21\times 10^{-25}\ M_P^4$, $a_{\rm reh}/a_{\rm end}\simeq 6.1\times 10^4$ (no preheating) and $a_{\rm reh}/a_{\rm end}\simeq 3.7\times 10^4$ ($\chi$ preheating). Numerically, we find $\rho_{\rm reh}\simeq 1.26\times 10^{-25}\ M_P^4$, $a_{\rm reh}/a_{\rm end}\simeq 4.3\times 10^4$ (no preheating) and $a_{\rm reh}/a_{\rm end}\simeq 2.6\times 10^4$ ($\chi$ preheating). The analytical estimates for $a_{\rm reh}$ are shifted by a factor of $\sim1.4$ since we have neglected the exponential decay of $\rho_{\phi}$ at the end of reheating~\cite{Turner:1983he,Ellis:2015pla}. Nevertheless, there is a few percent level agreement for $\rho_{\rm reh}$ and a sub-percent agreement for the shift factor $\Delta^{1/3}\simeq 0.60$ for $a_{\rm reh}$ in the presence of resonant DM production. 

\begin{figure}[!t]
    \includegraphics[width=\textwidth]{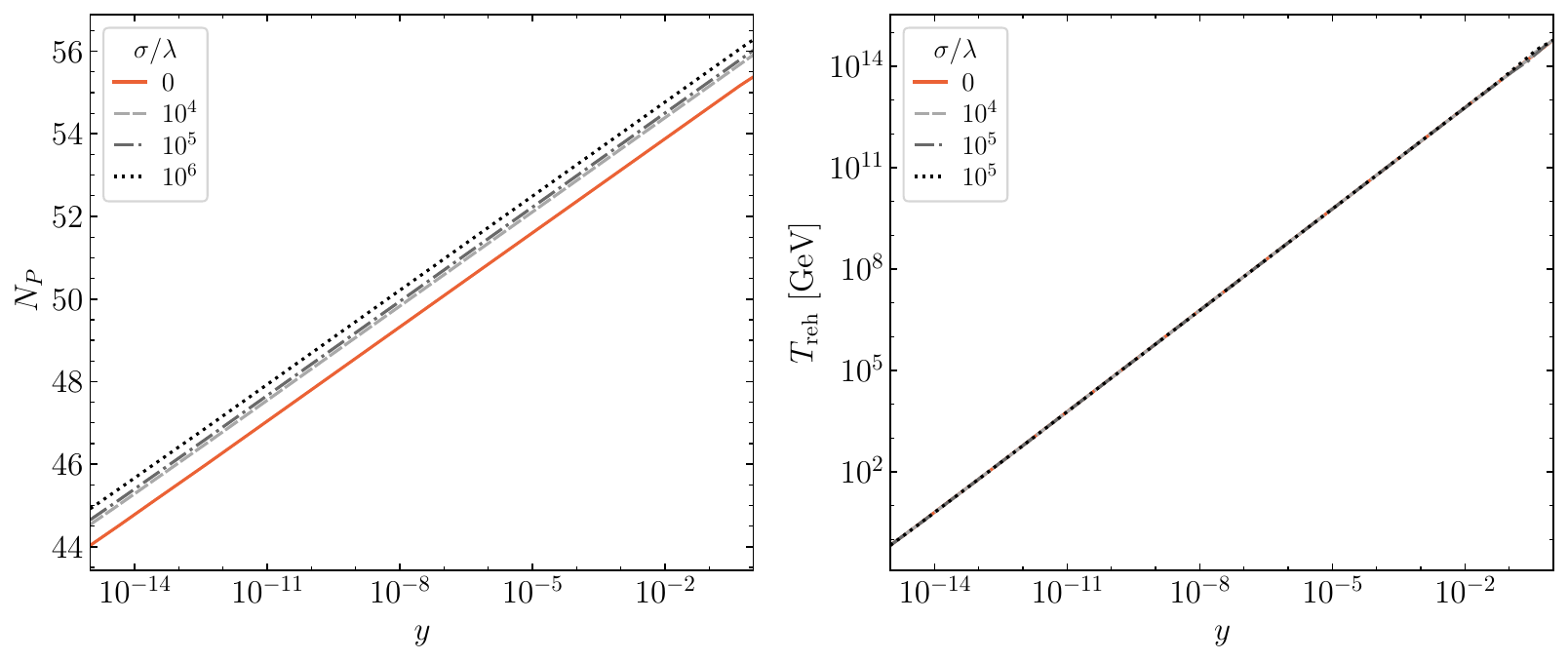}
    \caption{Number of $e$-folds at the {\em Planck} pivot scale (left) and reheating temperature (right), as a functions of the inflaton-matter Yukawa coupling $y$, for a selection of inflaton-DM effective couplings $\sigma/\lambda$.}
    \label{fig:NT}
\end{figure}

With the duration of reheating, and the dilution of the energy density, we can now evaluate the shift in the number of $e$-folds for a given pivot scale $k_*$. Substitution of (\ref{eq:arehpreh}) in (\ref{eq:Nstar}), with $a_{\rm rad}\propto a_{\rm reh}$, gives
\beq\label{eq:deltaN}
N_* - N_*^{(\rm pert)} \;\simeq\; - \frac{1}{3}\ln\Delta \;\simeq\; \begin{cases}
0.51\,,\ & \sigma/\lambda=10^4\,,\\
0.63\,,\ & \sigma/\lambda=10^5\,,\\
0.90\,,\ & \sigma/\lambda=10^6\,,
\end{cases}
\eeq
in excellent agreement with numerical results, shown in the first panel of Fig.~\ref{fig:NT}. The second panel of the figure demonstrates the independence of $T_{\rm reh}$ with respect to dark sector preheating, for perturbative Yukawa couplings. It is worth noting that Eq.~(\ref{eq:deltaN}) has the same functional dependence for the shift in $N_*$ in the case of a late-time injection of entropy by a factor $\Delta$, as is the case for embeddings of Starobinsky inflation in flipped SU(5) grand unified theories~\cite{Ellis:2018moe,Ellis:2021kad}.

\subsection{Dark matter relic abundance}\label{sec:DM}

In the absence of DM interactions other than its coupling to $\phi$, the number density produced during reheating is only redshifted to become the present DM relic abundance. Assuming the $\chi$ field is nonrelativistic at the present time, the dark matter relic abundance can be computed as
\begin{align} \label{eq:Omegachi1}
\Omega_{\chi} \;&=\; \frac{m_{\chi}n_{\chi}}{\rho_c} \;=\; \frac{m_{\chi} \, n_{\chi}(a_m)(a_m/a_{\rm end})^3}{\rho_c}\left(\frac{a_{\rm end}}{a_0}\right)^3\,,
\end{align}
where $\rho_c = 3H_0^2 M_P^2$ is the present critical density of the universe, and $n_{\chi}(a_m)(a_m/a_{\rm end})^3$ is the comoving DM number density, evaluated after backreaction and therefore constant in time. Following a standard computation, the amount of expansion between the end of inflation and today can be approximated as~\cite{Liddle:2003as,Martin:2010kz,Garcia:2022vwm}
\begin{align} \notag
\frac{a_{\rm end}}{a_0} \;&=\; \frac{a_{\rm end}}{a_{\rm reh}}\left(\frac{\rho_{\rm end}}{\rho_{\rm reh}}\right)^{1/4} \frac{H_0}{\rho_{\rm end}^{1/4}}\left(\frac{\rho_{\rm reh}^{1/4}a_{\rm reh}}{a_0H_0}\right)\\ \label{eq:aratio1}
&\simeq\; \Delta^{-1/3} \left( \frac{a_{\rm end}}{a_{\rm reh}} \right)_{\rm pert}\left(\frac{\rho_{\rm end}}{\rho_{\rm reh}}\right)^{1/4}_{\rm pert} \left(\frac{\pi^2}{90}\right)^{1/4}\left(\frac{43}{11}\right)^{1/3}g_{\rm reh}^{-1/12}\frac{T_0}{(H_{\rm end}M_P)^{1/2}}\\ \label{eq:aratio2}
&\simeq\; \Delta^{-1/3}\left(\frac{43\pi^2}{198\sqrt{10}}\right)^{1/3}\frac{T_{\rm reh}^{1/3}T_0}{(H_{\rm end}M_P)^{2/3}}\,,
\end{align}
where in arriving to (\ref{eq:aratio1}) we have used the approximation (\ref{eq:arehpreh}), and for (\ref{eq:aratio2}) we have made the approximation rad $\simeq$ reh in (\ref{eq:Rradpert}). Substitution into Eq.~(\ref{eq:Omegachi1}), including (\ref{eq:mphi}), then yields
\begin{align}
\Omega_{\chi}h^2 \;&\simeq\; \frac{43\pi^2 T_0^3 m_{\chi}T_{\rm reh} n_{\chi}(a_m)(a_m/a_{\rm end})^3 }{594\sqrt{10} (H_0 h^{-1})^2 H_{\rm end}^2 M_P^4 \Delta}\\ \label{eq:Omegachi2}
&\simeq\; \frac{0.12}{\Delta}\,\left(\frac{3\times 10^{13}\,{\rm GeV}}{m_{\phi}}\right)^2 \left(\frac{ n_{\chi}(a_m)(a_m/a_{\rm end})^3 }{1.6\times 10^{-12} M_P^3}\right)\left(\frac{m_{\chi}}{1\,{\rm GeV}}\right)\left(\frac{T_{\rm reh}}{10^{10}\,{\rm GeV}}\right)\left(\frac{N_*}{55}\right)^2\,.
\end{align}
This expression for the primordial DM closure fraction is an improvement over the computation presented previously in~\cite{Garcia:2022vwm}, as it takes into account the more prompt reheating due to DM backreaction.

\begin{figure}[!t]
\centering
    \includegraphics[width=0.80\textwidth]{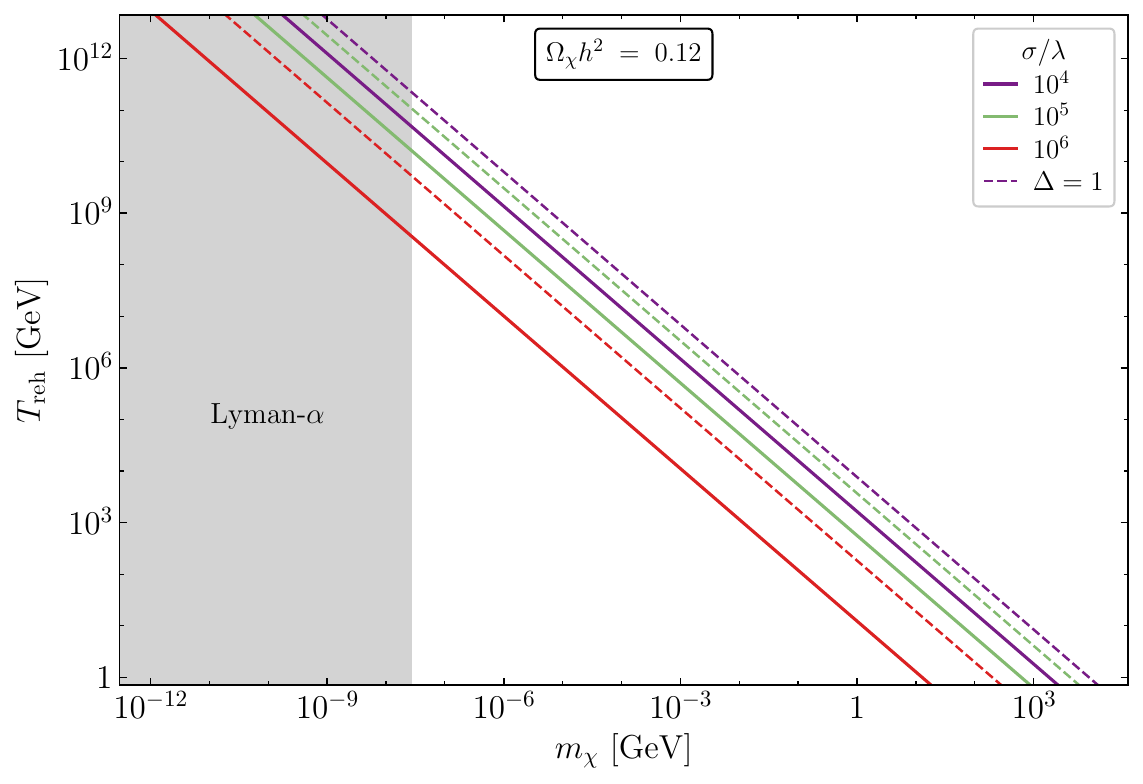}
    \caption{Parameter space for dark matter production from strong preheating, for a selection of effective couplings $\sigma/\lambda$. Along each line the measured dark matter relic abundance is $\Omega_{\chi}h^2=0.12$. For the dashed curves we ignore the change in the reheating duration due to DM backreaction. Along the continuous curves we use the full approximation (\ref{eq:Omegachi2}) with $\Delta\neq1$. The shaded region is excluded by the Lyman-$\alpha$ measurement of the matter power spectrum. }
    \label{fig:TvsM}
\end{figure}

Fig.~\ref{fig:TvsM} shows the reheating temperature necessary to saturate the observed value of $\Omega_{\chi}$, as a function of the DM mass. The continuous lines show the (inverse) relation accounting for the shift in $a_{\rm reh}$, $\Delta\neq 1$, while the dashed curves ignore this effect. As expected, the effect of $\Delta$ is to enhance the primordial abundance given a fixed $m_{\chi}$ and $T_{\rm reh}$. Therefore, lower masses/temperatures are required compared to the naive result with $\Delta=1$. This shift corresponds simply to a reduction by a factor of $\Delta$ in $T_{\rm reh}$ for a given $m_{\chi}$, or vice-versa, given that the logarithmic correction to $N_*$ is negligible in comparison (see Eq.~(\ref{eq:deltaN})). We note that high reheating temperatures require small (sub-keV) masses. In such cases, the $\chi$ field can behave at late times as non-cold dark matter, resulting in a suppression of clustered structure overdensities on galactic scales. This suppression is absent in Lyman-$\alpha$ measurements of the matter power spectrum above scales $k\simeq 15h\ {\rm Mpc}^{-1}$. For warm dark matter candidates, this constraint can be translated to a lower bound in the mass, $m_{\rm WDM}=(1.9-5.3)\,{\rm keV}$~\cite{Narayanan:2000tp,Viel:2005qj,Viel:2013fqw,Baur:2015jsy,Irsic:2017ixq,Palanque-Delabrouille:2019iyz,Garzilli:2019qki}. In the present case of non-thermal, resonant DM production, this lower bound can be translated upon matching the equation-of-state parameter of $\chi$ with the equation-of-state parameter of a thermal candidate with mass $m_{\rm WDM}$. The full details of this procedure can be found in~\cite{Ballesteros:2020adh}. In Refs.~\cite{Garcia:2022vwm,Garcia:2023awt} this mapping has been applied to the model at hand (\ref{eq:action}). The result is a quasi-universal lower bound in the backreaction regime, of $m_{\chi}\gtrsim 30\,{\rm eV}$. The shaded gray region in Fig.~\ref{fig:TvsM} corresponds to the parameter space ruled out by structure formation considerations. It is worth noting that the shortened reheating due to backreaction lowers the allowed range of temperatures. The maximum reheating temperatures correspond to
\beq
T_{{\rm reh,Ly}\mbox{-}\alpha} \;\simeq\; \begin{cases}
4.5\times 10^{10}\,{\rm GeV}\,,\ & \sigma/\lambda=10^4\,,\\
1.5\times 10^{10}\,{\rm GeV}\,,\ & \sigma/\lambda=10^5\,,\\
3.4\times 10^8\,{\rm GeV}\,,\ & \sigma/\lambda=10^6\,.
\end{cases}
\eeq
For more details on the phenomenology of DM production from preheating, including relic abundance, structure formation, isocurvature and reheating constraints, in the weak and strong coupling regimes, see~\cite{Garcia:2022vwm,Garcia:2023awt}. 

\subsection{CMB observables}\label{sec:CMB}

\begin{figure}[!t]
    \includegraphics[width=\textwidth]{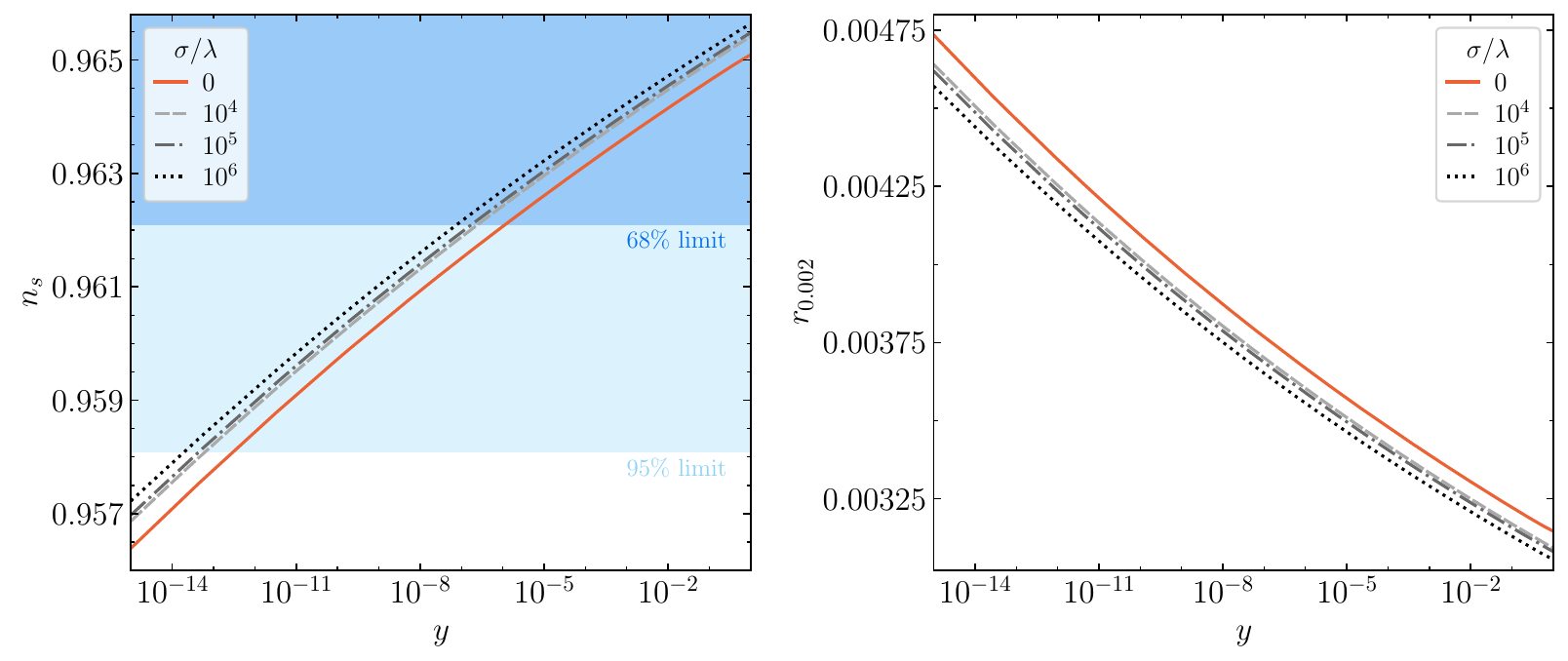}
    \caption{Scalar tilt (left) and tensor-to-scalar ratio (right), as a functions of the inflaton-matter Yukawa coupling $y$, for a selection of inflaton-DM effective couplings $\sigma/\lambda$. The left panel shows the 68\% and 95\% CL regions from {\em Planck} in combination with BK18+BAO~\cite{BICEP:2021xfz}. In the right panel these contours are not shown, since all curves lie inside the $1\sigma$ region. }
    \label{fig:nsrpreh}
\end{figure}

After having determined the effect of DM preheating on the number of $e$-folds after horizon crossing, we are now in position to compute the effect in the CMB observables $n_s$ and $r$. A straigthforward substitution of the numerical results in Fig.~\ref{fig:NT} onto the exact spectral parameters in Fig.~\ref{fig:nsr} produces the curves seen in Fig.~\ref{fig:nsrpreh}. The left panel shows the dependence of the spectral tilt on the Yukawa coupling $y$ in the absence of preheating (solid, red), and for the DM-inflaton couplings $\sigma/\lambda=\{10^4,10^5,10^6\}$ (dashed, dot-dashed and dotted, respectively). In this same panel, for reference, we show the 1$\sigma$ and 2$\sigma$ {\em Planck}+BK18+BAO CL contours, marginalized at $r_{\rm 0.002}\simeq 0.004$~\cite{BICEP:2021xfz}. As it is clear, the result is an increase in the tilt of the scalar power spectrum, larger for stronger $\phi$-$\chi$ couplings. This shift in $n_s$ is present at all $y< 1$. Combining expressions (\ref{eq:nsN}) and (\ref{eq:deltaN}), and the correction to the slow-roll prediction shown in Fig.~\ref{fig:nsr}, we can parametrize the effect of DM preheating on $n_s$ as
\beq
n_s \;\simeq\; n_s^{({\rm pert})} - \frac{2\ln\Delta}{3(N_*+2.5)^2}\,.
\eeq
This result is in excellent agreement with the numerical data, indistinguishable from it at the sub-percent level. The effect on $n_s$ from non-linear dynamics is therefore always small, $\delta n_s\lesssim 10^{-3}$. Such a change is smaller than the forecasted sensitivity of future CMB experiments such as CMB-S4~\cite{Abazajian:2019eic}, the Simons Observatory~\cite{SimonsObservatory:2018koc} or LiteBIRD~\cite{Hazumi:2019lys,LiteBIRD:2020khw}, although it could in principle be probed by EUCLID~\cite{EUCLID:2011zbd} or SKA~\cite{Maartens:2015mra,Sprenger:2018tdb}. Nevertheless, the shift in $n_s$ is important for constraining the allowed inflaton-SM couplings. Specifically, we find that the minimum coupling allowed by observations can differ by more than an order of magnitude at the 68\% and 95\% CL, as shown in Table~\ref{tab:nslims}. A similar result applies for the minimum reheating temperature, also shown in Table~\ref{tab:nslims}. At $\sigma/\lambda=0$ the bounds presented here are consistent with those in~\cite{Ellis:2021kad}, accounting for a difference in the number of degrees of freedom $g_{\rm reh}$ at high temperatures (SM vs MSSM), and a slight modification in the numerical method.\par\medskip

\begin{table}[!t]
\centering
 \begin{tabular}{c | c c | c c } 
 \midrule
 $\sigma/\lambda$ & $y_{68\%}$ & $T_{\rm reh,68\%}$ [GeV] & $y_{95\%}$ & $T_{\rm reh,95\%}$ [GeV]  \\
 \midrule
 0 & $1.7\times 10^{-6}$ & $6.3\times10^8$ & $3.4\times 10^{-13}$ & $203$\\[4pt]
 $10^4$ & $2.5\times 10^{-7}$ & $1.5\times10^8$ & $6.9\times 10^{-14}$ & $42$\\[4pt]
 $10^5$ & $1.7\times 10^{-7}$ & $1.0\times10^8$ & $4.8\times 10^{-14}$ & $29$\\[4pt]
 $10^6$ & $8.3\times 10^{-8}$ & $4.9\times10^7$ & $1.9\times 10^{-14}$ & $12$\\
 \midrule
 \end{tabular}
 \caption{Lower bounds on the inflaton-SM Yukawa coupling $y$ and the reheating temperature $T_{\rm reh}$, as functions of the effective coupling $\sigma/\lambda$, at the 68\% and 95\% CL. These constraints are particular to the Starobinsky model (\ref{eq:staropot}), derived from the marginalized {\em Planck}+BK18+BAO bounds on $n_s$ at $r_{0.002}\simeq 0.004$. }
\label{tab:nslims}
\end{table}

The right panel of Fig.~\ref{fig:nsrpreh} shows the numerically computed values for the tensor-to-scalar ratio $r$ as a function of the Yukawa coupling, in the absence and presence of preheating, analogously to the $n_s$ panel. Here we do not depict the {\em Planck}+BICEP/{\em Keck} constraints, as for all couplings the tensor-to-scalar ratio is below the $1\sigma$ upper bound $r\lesssim 0.035$~\cite{BICEP:2021xfz}. We note in this case that the effect of preheating is a reduction in the tensor amplitude relative to the scalar spectrum. This effect can be justified upon combining the slow-roll approximation (\ref{eq:rN}), together with (\ref{eq:deltaN}) and the shift $N_*\rightarrow N_*+2.5$ which corrects the slow-roll result (see Fig.~\ref{fig:nsr}),
\beq
r \;\simeq\; r^{({\rm pert})} + \frac{8\ln\Delta}{(N_*+2.5)^3}\,.
\eeq
Similarly to $n_s$, the effect on $r$ is far below the sensitivity threshold of the next generation of CMB observatories. Moreover, the constraints on the inflaton couplings to other fields are largely independent of $r$ for small values of it. Therefore, the effect of preheating on gravitational waves sourced around CMB frequencies can be neglected for plateau-like potentials. Nevertheless, as is well known, the non-linear growth of scalar inhomogeneities during preheating can efficiently source gravitational waves at high frequencies, $f\sim10^8$ Hz~\cite{Khlebnikov:1997di,Easther:2006gt,Easther:2006vd,Garcia-Bellido:2007nns,Garcia-Bellido:2007fiu,Dufaux:2007pt,Figueroa:2017vfa}. At CMB scales, the main effect would correspond to an enhancement of the effective number of relativistic species, $\Delta N_{\rm eff}=N_{\rm eff}-N_{\rm eff}^{\rm SM}$, with $N_{\rm eff}^{\rm SM}=3.046$. However, the resulting enhancement is typically $\Delta N_{\rm eff}\lesssim 10^{-4}$~\cite{Figueroa:2017vfa,Cosme:2022htl}, well below the expected precision at the $2\sigma$ level of CMB-S4 ($\Delta N_{\rm eff}<0.06$)~\cite{Abazajian:2019eic}, CMB-HD ($\Delta N_{\rm eff}<0.027$)~\cite{CMB-HD:2022bsz}, or COrE and Euclid ($\Delta N_{\rm eff}<0.013$)~\cite{COrE:2011bfs,EUCLID:2011zbd}.

\section{Summary} \label{sec:conclusions}

One of the largest theoretical uncertainties in mapping the predictions of any inflationary model to measurements of the primordial power spectra is the precise characterization of reheating. Under the assumption of perturbative inflaton decays, the parametrization of the duration of reheating, and the amount of dilution of the energy density, encoded in the reheating parameter $R_{\rm rad}$, can be simply determined in terms of a single effective inflaton-matter coupling, often parametrized in terms of the reheating temperature $T_{\rm reh}$~\cite{Martin:2010kz,Mielczarek:2010ag,Dai:2014jja,Martin:2014nya,Cook:2015vqa,Ellis:2015pla,Drewes:2015coa,Ellis:2021kad,Drewes:2022nhu,Drewes:2023bbs}. The result is a straightforward connection between the main CMB observables $n_s$ and $r$, and the microphysical inflaton couplings to other fields (c.f.~Eq.~(\ref{eq:Rradpert})). However, when particle production is enhanced (or suppressed) by collective phenomena, the perturbative approximation fails, and a simple analytical description of reheating dynamics is not available. This in turn prevents a simple relation between observables and field couplings. 

In this paper we have determined the impact of resonant dark matter production during reheating on the evaluation of CMB observables. Since generality is broken in the presence of nonlinearity, we have limited our numerical study to the Starobinsky model of inflation, a scalar dark sector, and perturbative inflaton decays into Standard Model particles. Specifically, we have considered the four-legged interaction $\mathcal{L}\supset - \sigma \phi^2\chi^2/2$ between the inflaton $\phi$ and the DM $\chi$, whose strength can be characterized by the dimensionless ratio $\sigma/\lambda$, with $\lambda = m_{\phi}^2/2M_P^2$. For $\sigma/\lambda\gtrsim 5\times 10^3$~\cite{Garcia:2022vwm}, parametric resonance drives an exponential growth of the DM mode functions, into the backreaction regime. These nonlinear dynamics not only result in a reduction of the comoving energy density of the inflaton by an $\mathcal{O}(10)$ factor, but also in the conversion of the remaining coherent, homogenous inflaton field into free inflaton quanta. The efficiency of the decay of the inflaton into SM particles is in consequence reduced, although the transition time from reheating to radiation domination is also reduced. 

In order to map the aforementioned shift on the duration of reheating to CMB observables, we have accurately tracked the energy and number densities of the DM field and the inflaton throughout the backreaction regime, with the help of \CL~\cite{Figueroa:2020rrl,Figueroa:2021yhd}. With them, we computed the perturbative SM particle production rate, following the evolution of the energy density of the primordial plasma before, during and after the transient period of preheating. We have found that reheating temperatures are impervious to dark sector preheating, while the duration of reheating, and the number of $e$-folds $N_*$, can be simply parametrized in terms of the loss in comoving energy density of the inflaton during backreaction, denoted by $\Delta$. This has allowed us to find a functional form for the correction to the CMB observables, $n_s(\Delta)$, $r(\Delta)$. Although we find that the resulting increase in $n_s$ and the decrease in $r$ are too small to be comparable to the sensitivity of next-generation CMB experiments, they modify the favored range of inflaton-SM couplings by up to an order of magnitude. In terms of the reheating temperature, $T_{\rm reh}<203$ GeV are excluded at the 95\% CL in the absence of preheating, while in the presence of strong preheating $T_{\rm reh}$ can be as low as $12$ GeV. 

Additionally, we have revisited in this work the calculation of the primordial DM relic abundance produced from the parametric excitation of the dark scalar field, previously explored in~\cite{Garcia:2022vwm,Garcia:2023awt}. We have improved the calculation of the DM closure fraction by including the shift in the duration of reheating due to the depletion of the comoving inflaton energy density due to DM preheating. We find that the shortened duration of reheating enhances the relic abundance as $\Omega_{\chi}\rightarrow \Omega_{\chi}/\Delta$. The allowed range of reheating temperatures for a given inflaton-DM coupling and DM mass is therefore reduced by the same factor. Table~\ref{tab:summary} summarizes the constraints on $T_{\rm reh}$ and $m_{\chi}$ from structure formation and CMB considerations. Our first conclusion is that only light DM masses are allowed, $\lesssim 40$ GeV at the $1\sigma$ level. Our second and main conclusion is that, at the CMB 68\% CL, only a narrow range of reheating temperatures is allowed. In particular, for strong preheating with $\sigma/\lambda=10^6$, the allowed range for $T_{\rm reh}$ spans only one order of magnitude. We emphasize that these results are applicable only for resonant DM production in Starobinsky inflation. Nevertheless, as is discussed in~\cite{Garcia:2023awt}, we expect our conclusions to be qualitatively unchanged for plateau-like inflationary potentials. Moreover, for alternative realizations of inflation and dark sector preheating, we have provided here a roadmap to accurately constrain the inflaton couplings. Such constraints on the microphysics of inflation are crucial for narrowing down the UV completions of the Standard Model within which the inflationary paradigm must be embedded.

\begin{table}[!t]
\centering
 \begin{tabular}{c | r c c c l l } 
 \midrule
 $\sigma/\lambda$ & Lyman-$\alpha$ & & & & CMB 68\% & CMB 95\%  \\ 
 \midrule
 $10^4$ & $4.5\times 10^{10}$ GeV & $\gtrsim$ & $T_{\rm reh}$ & $\gtrsim$ & $1.5\times10^8$ GeV & ($42$ GeV)\\
  & $30$ eV & $\lesssim$ & $m_{\chi}$ & $\lesssim$ & $9$ keV & ($40$ GeV)\\[6pt]
 $10^5$ & $1.5\times 10^{10}$ GeV & $\gtrsim$ & $T_{\rm reh}$ & $\gtrsim$ & $1.0\times10^8$ GeV & ($29$ GeV)\\
  & $30$ eV & $\lesssim$ & $m_{\chi}$ & $\lesssim$ & $5$ keV & ($20$ GeV)\\[6pt]
 $10^6$ & $3.4\times 10^{8}$ GeV & $\gtrsim$ & $T_{\rm reh}$ & $\gtrsim$ & $4.9\times10^7$ GeV & ($12$ GeV)\\
  & $30$ eV & $\lesssim$ & $m_{\chi}$ & $\lesssim$ & $0.2$ keV & ($1$ GeV)\\
 \midrule
 \end{tabular}
 \caption{Summary table for the upper and lower bounds on the reheating temperature and dark matter mass for a selection of effective inflaton-DM couplings $\sigma/\lambda$, from structure formation constraints (upper bound) and the 68\% and 95\% CL {\em Planck}+BK18 bounds on $n_s$ for Starobinsky inflation (lower bounds). }
\label{tab:summary}
\end{table}

\begin{acknowledgments}
We would like to thank Mathias Pierre and Sarunas Verner for helpful discussions. MG and AP are supported by the DGAPA-PAPIIT grant IA103123 at
UNAM, and by the CONAHCYT ``Ciencia de Frontera'' grant CF-2023-I-17. We acknowledge Instituto de F\'isica (UNAM) for providing computing infrastructure, and
Carlos L\'opez Natar\'en for his support in the use of the Holiday cluster. Non-perturbative numerical results in the Hartree approximation were obtained from a custom Fortran code utilizing the thread-safe arbitrary precision package MPFUN-For~\cite{mpfun}.
\end{acknowledgments}

\addcontentsline{toc}{section}{References}
\bibliographystyle{utphys}
\bibliography{references} 

\end{document}